\documentclass[journal,12pt,onecolumn]{IEEEtranTCOM}

\usepackage{setspace}
\ifCLASSOPTIONonecolumn \doublespace \fi
\usepackage{graphics}
\usepackage{rotating}
\usepackage{amsfonts}
\usepackage{amsmath}
\usepackage{amsthm}
\usepackage{subfigure}
\usepackage{multirow}
\allowdisplaybreaks[1]
\newtheorem*{theorem}{Theorem}
\renewcommand{\IEEEQED}{\IEEEQEDopen}

\usepackage{cite}

\graphicspath{{Figures/}} \DeclareGraphicsExtensions{.eps,.ps}

\title{Full-Diversity Precoding Design of Bit-Interleaved Coded Multiple Beamforming with Orthogonal Frequency Division Multiplexing}

\ifCLASSOPTIONconference
\author{\IEEEauthorblockN{Boyu Li and Ender Ayanoglu}\\
\IEEEauthorblockA{Center for Pervasive Communications and Computing\\
Department of Electrical Engineering and Computer Science\\
University of California, Irvine\\
Email: boyul@uci.edu, ayanoglu@uci.edu}} 
\else
\author{Boyu~Li,~\IEEEmembership{Member,~IEEE,}~and~Ender~Ayanoglu,~\IEEEmembership{Fellow,~IEEE}
\thanks{B. Li and E. Ayanoglu are with the Center for Pervasive Communications and Computing, Department of Electrical Engineering and Computer Science, Henry Samueli School of Engineering, University of California, Irvine, CA 92697-3975 USA (e-mail: boyul@uci.edu; ayanoglu@uci.edu).}} \fi

\begin{document}
\maketitle


\begin{abstract}

Multi-Input Multi-Output (MIMO) techniques have been incorporated with Orthogonal Frequency Division Multiplexing (OFDM) for broadband wireless communication systems. Bit-Interleaved Coded Multiple Beamforming (BICMB) can achieve both spatial diversity and spatial multiplexing for flat fading MIMO channels. For frequency selective fading MIMO channels, BICMB with OFDM (BICMB-OFDM) can be employed to provide both spatial diversity and multipath diversity, making it an important technique. In our previous work, the subcarrier grouping technique was applied to combat the negative effect of subcarrier correlation. It was also proved that full diversity of BICMB-OFDM with Subcarrier Grouping (BICMB-OFDM-SG) can be achieved within the condition $R_cSL \leq 1$, where $R_c$, $S$, and $L$ are the code rate, the number of parallel streams at each subcarrier, and the number of channel taps, respectively. The full diversity condition implies that if $S$ increases, $R_c$ may have to decrease to maintain full diversity. As a result, increasing the number of parallel streams may not improve the total transmission rate. In this paper, the precoding technique is employed to overcome the full diversity restriction issue of $R_cSL \leq 1$ for BICMB-OFDM-SG. First, the diversity analysis of precoded BICMB-OFDM-SG is carried out. Then, the full-diversity precoding design is developed with the minimum achievable decoding complexity. 

\end{abstract}



\section{Introduction} \label{sec:Introduction}

The demand for wireless services is constantly increasing. At the same time, research on Multi-Input Multi-Output (MIMO) wireless systems has been ongoing for more than a decade, with increasing success. MIMO systems are a way to attack the increasing capacity demand for wireless services since they can offer superior spectral efficiency and improved performance within a given bandwidth. A commonly employed MIMO approach is known as beamforming via Singular Value Decomposition (SVD) of the MIMO channel matrix. This approach enables spatial multiplexing\footnotemark \footnotetext{In this paper, the term ``spatial multiplexing" is used to describe the number of spatial subchannels, as in \cite{Paulraj_ST}. Note that the term is different from ``spatial multiplexing gain" defined in \cite{Zheng_DM}.}, enabling increased data rates. It can also enhance system performance. This technique requires the Channel State
Information (CSI) to be available at both the transmitter and receiver \cite{Jafarkhani_STC}.

For flat fading MIMO channels, it has been shown that employing beamforming with only one spatial channel, or transmitting one symbol
at a time, achieves the full diversity order provided by the channel \cite{Sengul_DA_SMB, Ordoez_HSAP}. In addition, employing a MIMO beamformer with more than one spatial channel without any channel coding results in the loss of full diversity \cite{Sengul_DA_SMB, Ordoez_HSAP}. On the other hand, employing Bit Interleaved Coded Modulation (BICM) \cite{Caire_BICM} together with SVD beamforming restores full diversity \cite{Akay_FSPFD, Akay_BICMB, Akay_On_BICMB}. Such a system was analyzed and called Bit-Interleaved Coded Multiple Beamforming (BICMB) in \cite{Akay_FSPFD, Akay_BICMB, Akay_On_BICMB}. In the study of BICMB, so far, the channel coding technique employed has been convolutional codes \cite{Lin_ECC}. The output of the convolutional code is then interleaved through the multiple subchannels with different diversity orders. BICMB can achieve the full diversity order as long as the code rate $R_c$ and the number of spatial channels $S$ satisfy $R_c S \leq 1$ \cite{Park_DA_BICMB, Park_DA_BICMB_J}. It has been further shown that when a constellation precoding technique is used in uncoded and coded SVD beamforming, this latter condition can be overridden and full diversity and full spatial multiplexing offered by the channel can be simultaneously achieved \cite{Park_CPB, Park_BICMB_CP, Park_MB_CP, Park_CPMB}. Specifically, without channel coding, full diversity requires that all spatial channels are precoded. In the case of precoding with BICMB, partial precoding can achieve both full diversity and full spatial multiplexing. Partial precoding is desirable because it reduces the high complexity of decoding a fully precoded BICMB system. These precoders now result in a system that achieves full diversity even without satisfying the condition $R_cS \leq 1$. As an alternative to precoding as in \cite{Park_CPB, Park_BICMB_CP, Park_MB_CP, Park_CPMB}, Perfect Space-Time Block Codes (PSTBCs) \cite{Oggier_PSTBC} were employed as the precoding technique \cite{Li_GCMB, Li_BICMB_PC, Li_MB_PC}. PSTBCs have desirable properties of full rate, full diversity, uniform average transmitted energy per antenna, good shaping of the constellation, and nonvanishing constant minimum determinant for increasing spectral efficiency. The resulting system achieves almost the same performance as precoded BICMB while reducing the decoding complexity substantially, for MIMO dimensions $2$ and $4$ \cite{Li_GCMB, Li_BICMB_PC, Li_MB_PC}. 

If the fading in the channel is not flat, or when the channel has frequency selective fading, the result is Inter-Symbol Interference
(ISI) for the transmitted symbols. Orthogonal Frequency Division Multiplexing (OFDM) is commonly used to combat ISI caused by multipath propagation \cite{Barry_DC}. OFDM transmits data in a parallel fashion on closely spaced subcarriers. The subcarriers satisfy an orthogonality property in order to reduce bandwidth. OFDM is robust against ISI. It achieves this by using equalization in the frequency domain with the advantage of avoiding the computational burden and the long convergence time requirements associated with time domain equalization. Therefore, OFDM can adapt to severe channel conditions. In addition, OFDM has high spectral efficiency, efficient implementation using Fast Fourier Transform (FFT) and Inverse FFT (IFFT), and low sensitivity to time synchronization errors.
With OFDM, multipath diversity can be achieved by adding channel coding \cite{Akay_BICM_OFDM, Akay_BICM_OFDM_STBC}. As a result, OFDM is well-suited for broadband data transmission, and it has been selected as the air interface for the Institute of Electrical and Electronics Engineers (IEEE) 802.11 Wireless Fidelity (WiFi) standard, the IEEE 802.16 Worldwide Interoperability for Microwave Access (WiMAX) standard, as well as the Third Generation Partnership Project (3GPP) Long Term Evolution (LTE) standard \cite{Ghosh_LTE}. 

The combination of MIMO and OFDM has been incorporated for all broadband wireless communication standards, i.e., WiFi \cite{IEEE_802_11}, WiMAX \cite{IEEE_802_16}, and LTE \cite{3GPP_TS_36.201}. For frequency selective MIMO channels, combining beamforming with OFDM can combat ISI and achieve spatial diversity \cite{Zamiri_MIMO_OFDM}. Moreover, both spatial diversity and multipath diversity can be achieved by adding channel coding, e.g., BICMB with OFDM (BICMB-OFDM), \cite{Akay_BICMB_OFDM, Akay_BICMB_OFDM_CSI, Akay_BICMB}. Although more sophisticated codes, such as turbo codes and Low-Density Parity-Check (LDPC) codes \cite{Lin_ECC}, are employed by some of the more modern standards \cite{Ghosh_LTE}, than convolutional codes employed by BICMB-OFDM, convolutional codes are still important because they can be analyzed and there are a great deal of legacy products using them. Therefore, BICMB-OFDM can be an important technique for broadband wireless communication. The diversity analysis of BICMB-OFDM was carried out in \cite{Li_DA_BICMB_OFDM}. In \cite{Li_DA_BICMB_OFDM}, the subcarrier grouping technique was employed to overcome the performance degradation caused by subcarrier correlation and offer multi-user compatibility. It was proved that full diversity of BICMB-OFDM with Subcarrier Grouping (BICMB-OFDM-SG) can be achieved as long as the condition $R_cSL \leq 1$ is satisfied, where $S$ is the number of streams transmitted at each subcarrier and $L$ is the number of channel taps. The full diversity condition implies that if the number of streams $S$ transmitted at each subcarrier increases, the code rate $R_c$ may have to decrease in order to keep full diversity. Hence, increasing the number of parallel streams may not improve the total transmission rate, which is a similar issue to the full diversity condition $R_cS \leq 1$ of BICMB for flat fading MIMO channels \cite{Park_DA_BICMB, Park_DA_BICMB_J}. Since precoding techniques have been successfully used to solve the full diversity restriction issue of $R_cS \leq 1$ for BICMB in the case of flat fading MIMO channels \cite{Park_CPB, Park_BICMB_CP, Park_MB_CP, Park_CPMB, Li_GCMB, Li_BICMB_PC, Li_MB_PC}, it may be possible to apply these techniques to BICMB-OFDM-SG so that its full diversity condition is not restricted to $R_cSL \leq 1$ for frequency selective MIMO channels. Nevertheless, the design criteria and diversity analysis cannot be generalized in a straightforward manner because of the increased system complexity.

In this paper, the main contribution is that the precoding technique is employed to solve the full diversity restriction issue of $R_cSL \leq 1$ for BICMB-OFDM-SG proposed in \cite{Li_DA_BICMB_OFDM}. First, diversity analysis of precoded BICMB-OFDM-SG is carried out. Based on the analysis, a full diversity condition related to the combination of the precoding matrix, the convolutional code, and the bit interleaver is provided. Then, the full diversity precoding design is developed. This design provides a sufficient method to guarantee full diversity while minimizing the increased decoding complexity caused by precoding.

The remainder of this paper is organized as follows: Section \ref{sec:basis} briefly describes BICMB-OFDM and BICMB-OFDM-SG. In Section \ref{sec:system}, the system model of BICMB-OFDM-SG employing precoding is proposed. In Section \ref{sec:diversity}, the diversity analysis of precoded BICMB-OFDM-SG is carried out. Then, Section \ref{sec:design} develops the full-diversity precoding design with the minimum achievable decoding complexity. In Section \ref{sec:results}, simulation results are provided. Finally, a conclusion is drawn in Section \ref{sec:conclusions}.

\section{Background Knowledge of BICMB-OFDM} \label{sec:basis}

In this section, a brief description of BICMB-OFDM is provided in order to offer sufficient background knowledge for the following sections. More details of BICMB-OFDM can be found in \cite{Li_DA_BICMB_OFDM}.

\subsection{BICMB-OFDM} \label{subsec:bicmb_ofdm}

\ifCLASSOPTIONonecolumn
\begin{figure}[!t]
\centering \includegraphics[width = 1.0\linewidth]{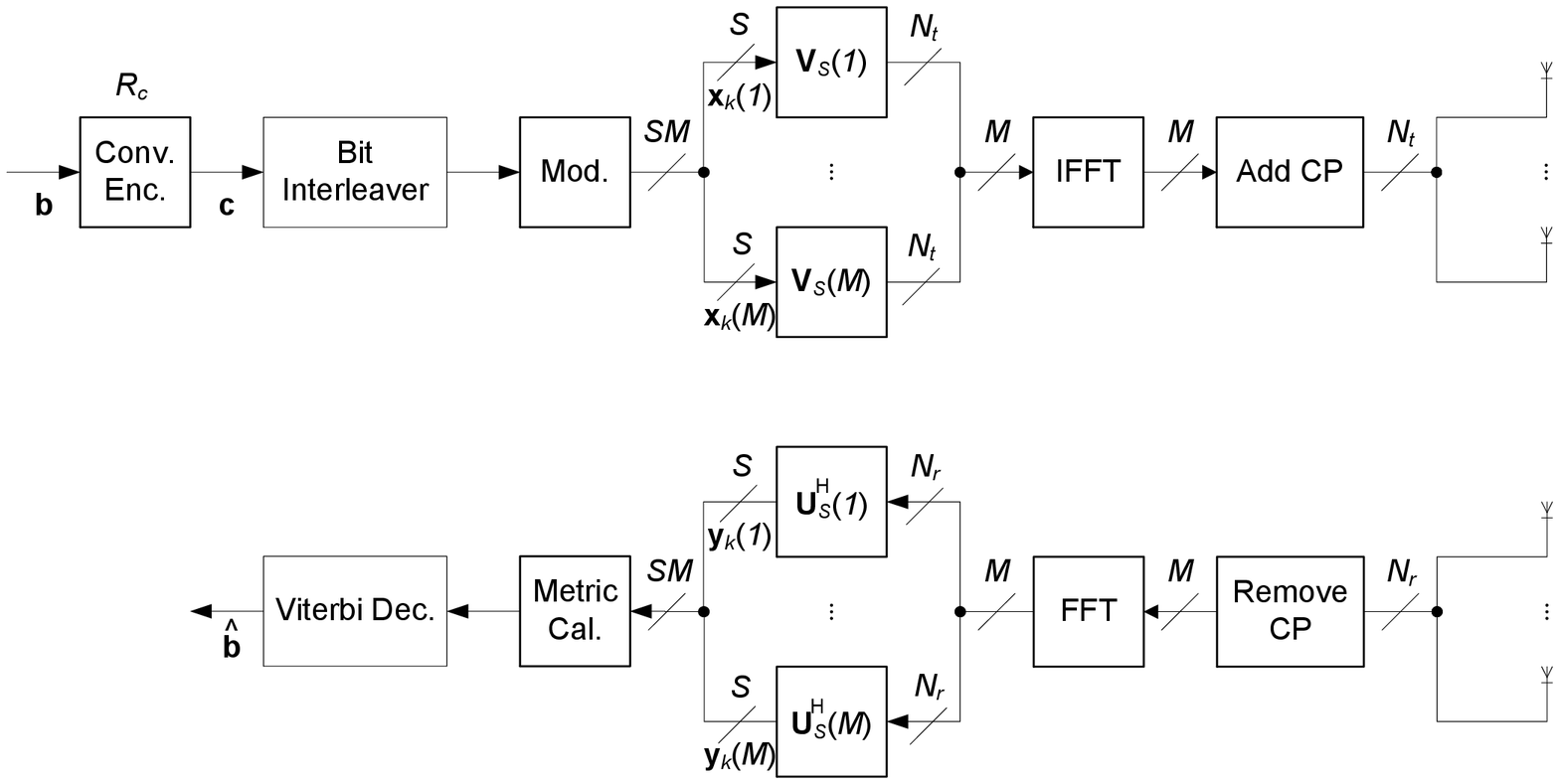}
\caption{Structure of BICMB-OFDM.} \label{fig:bicmb_ofdm}
\end{figure}
\else
\begin{figure}[!t]
\centering \includegraphics[width = 1.0\linewidth]{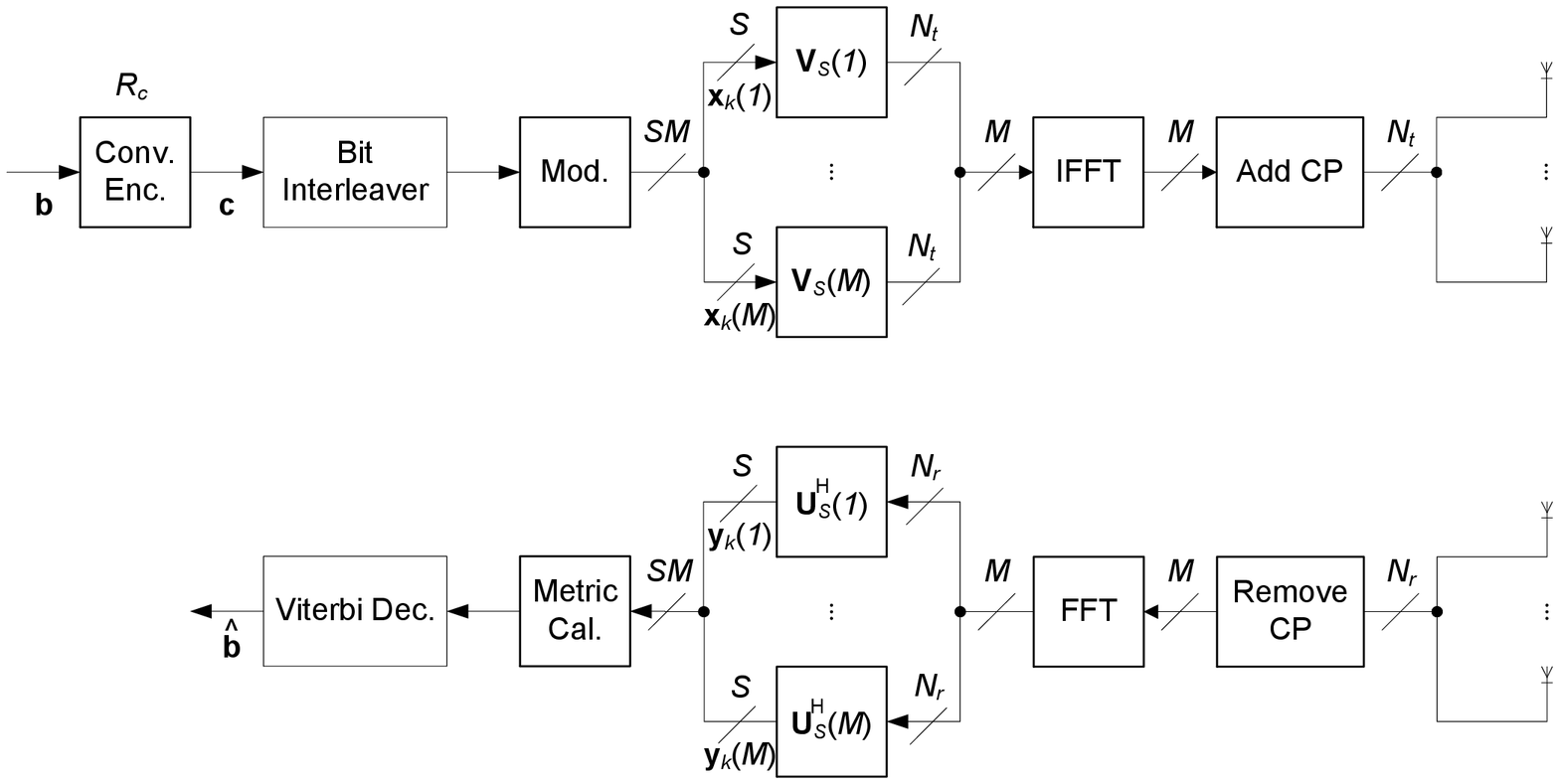}
\caption{Structure of BICMB-OFDM.} \label{fig:bicmb_ofdm}
\end{figure}
\fi

For frequency selective MIMO channels, BICMB-OFDM was proposed to provide both spatial diversity and multipath diversity \cite{Akay_BICMB_OFDM, Akay_BICMB_OFDM_CSI, Akay_BICMB}. Fig. \ref{fig:bicmb_ofdm} presents the structure of BICMB-OFDM. First, the bit codeword $\mathbf{c}$ is generated from the information bits by the convolutional encoder of code rate $R_c$, which is possibly combined with a perforation matrix for a high rate punctured code \cite{Haccoun_PCC}. After that, a random bit interleaver is applied to generate an interleaved bit sequence, which is then modulated, e.g., Quadrature Amplitude Modulation (QAM), to a symbol sequence. The number of transmit and receive antennas are denoted by $N_t$ and $N_r$ respectively. Assume that $M$ subcarriers are employed to transmit the symbol sequence, and $S \leq \min\{N_t,N_r\}$ parallel streams realized by SVD in the frequency domain for each subcarrier are transmitted at the same time. Hence, an $S \times 1$ symbol vector $\mathbf{x}_k(m)$ is carried on the $m$th subcarrier at the $k$th time instant with $m = 1,\ldots,M$. The length of Cyclic Prefix (CP), which is employed for OFDM to combat ISI caused by multipath propagation, is assumed to be $L_{cp}$ where $L_{cp} \geq L$ with $L$ denoting the number of channel taps.

The $L$-tap frequency selective fading MIMO channel is assumed to be Rayleigh quasi-static and known by both the transmitter and receiver, which is denoted by $\breve{\mathbf{H}}(l) \in \mathbb{C}^{N_r \times N_t}$ with $l=1,\ldots,L$ where $\mathbb{C}$ stands for the set of complex numbers. Let 
\begin{align}
\mathbf{H}(m)=\sum_{l=1}^{L} \breve{\mathbf{H}}(l)\exp\left(-i{2\pi (m-1) \tau_l \over MT}\right) \label{eq:channel_frequency}
\end{align}
denote the quasi-static flat fading MIMO channel observed at the $m$th subcarrier, where $T$ denotes the sampling period, $\tau_l$ indicates the $l$th tap delay, and $i=\sqrt{-1}$ \cite{Lee_ST_BICM_OFDM}. Then, the beamforming matrices at the $m$th subcarrier are determined by SVD of $\mathbf{H}(m)$, i.e., $\mathbf{H}(m) = \mathbf{U}(m) \mathbf{\Lambda}(m) \mathbf{V}^H(m)$, where the $N_r \times N_r$ matrix $\mathbf{U}(m)$ and the $N_t \times N_t$ matrix $\mathbf{V}(m)$ are unitary, and the $N_r \times N_t$ matrix $\mathbf{\Lambda}(m)$ is a rectangular diagonal matrix whose $s$th diagonal element, $\lambda_s(m) \in \mathbb{R}^+$, is a singular value of $\mathbf{H}(m)$ in decreasing order with $s=1,\ldots,S$ where $\mathbb{R}^+$ denotes the set of positive real numbers. When $S$ streams are transmitted for each subcarrier at the same time, the first $S$ columns of $\mathbf{U}(m)$ and $\mathbf{V}(m)$, i.e., $\mathbf{U}_S(m)$ and $\mathbf{V}_S(m)$, are chosen as beamforming matrices at the receiver and transmitter for the $m$th subcarrier respectively.

For each subcarrier, the multiplications with beamforming matrices $\mathbf{V}_S(m)$ and $\mathbf{U}_S^H(m)$ are carried out at each subcarrier before executing IFFT and adding CP at the transmitter, and after executing FFT and removing CP at the receiver, respectively. Therefore, the input-output relation of BICMB-OFDM for the $m$th subcarrier at the $k$th time instant is
\begin{align}
y_{s,k}(m) = {\lambda}_s(m) x_{s,k}(m) + n_{s,k}(m), \label{eq:detected_symbol}
\end{align}
for $s=1,\ldots,S$, where $y_{s,k}(m)$ and $x_{s,k}(m)$ are the $s$th element of the $S \times 1$ received symbol vector $\mathbf{y}_k(m)$ and the transmitted symbol vector $\mathbf{x}_k(m)$ respectively, and $n_{s,k}(m)$ is the additive white Gaussian noise with zero mean and variance $N_0=N_t/\gamma$ \cite{Liu_STF_OFDM} with $\gamma$ denoting the received Signal-to-Noise Ratio (SNR) over all the receive antennas. Note that the total transmitted power is scaled by $N_t$ in order to make the received SNR $\gamma$.

The location of the coded bit $c_{k'}$ within the transmitted symbol is denoted as $k' \rightarrow (k, m, s, j)$, meaning that the coded bit $c_{k'}$ is mapped onto the $j$th bit position on the label of $x_{s,k}(m)$. Let $\chi$ denote the signal set of the modulation scheme, and let $\chi_b^j$ denote a subset of $\chi$ whose labels have $b \in \{0, 1\}$ at the $j$th bit position. By using the location information $k' \rightarrow (k, m, s, j)$ and the input-output relation in (\ref{eq:detected_symbol}), the receiver calculates the Maximum Likelihood (ML) bit metrics for $c_{k'}=b$ as
\begin{align}
\Delta(y_{s,k}(m), c_{k'}) = \min_{x \in \chi_{c_{k'}}^j} \left| y_{s,k}(m) - {\lambda}_s(m)x \right|^2. \label{eq:ml_bit_metrics}
\end{align}

Finally, the ML decoder, which applies the soft-input Viterbi decoding \cite{Lin_ECC} to find a codeword with the minimum sum weight, makes decisions based on the rule given by \cite{Caire_BICM} as
\begin{align}
\mathbf{\hat{c}} = \arg\min_{\mathbf{c}} \sum_{k'} \Delta(y_{s,k}(m), c_{k'}).
\label{eq:decision_rule}
\end{align}

In \cite{Li_DA_BICMB_OFDM}, the diversity analysis of BICMB-OFDM was carried out. According to the analysis, the maximum achievable diversity of BICMB-OFDM was derived and the full diversity restriction of $R_cSL \leq 1$ was proved. In addition, the performance degradation due to subcarrier correlation was investigated, which showed that although the maximum achievable diversity is the same when SNR is relatively high, strong subcarrier correlation can result in significant performance loss for SNRs in the  practical range. 

\subsection{Subcarrier Grouping} \label{subsec:grouping}

In order to combat the performance degradation of BICMB-OFDM caused by subcarrier correlation, the subcarrier grouping technique was employed in \cite{Li_DA_BICMB_OFDM}. Instead of transmitting one stream of information through all subcarriers of OFDM, subcarrier grouping technique transmits multiple streams of information through multiple group of subcarriers, which was also suggested for multi-user interference elimination \cite{Wang_WMC}, Peak-to-Average Ratio (PAR) reduction \cite{Goeckel_OFDM}, and complexity reduction \cite{Liu_LCP_OFDM}. For BICMB-OFDM-SG, assuming that $G=M/L \in \mathbb{Z}$ where $\mathbb{Z}$ denotes the set of integer numbers, then $G$ streams of bit codewords are carried on $G$ different groups of $L$ uncorrelated or weakly correlated subcarriers at the same time.
\\\textbf{Example:}
Consider the case of $L=2$ and $M=64$. Then, the $g$th and the $(g+32)$th subcarriers are uncorrelated or weakly correlated for $g=1,\ldots,32$. Then, the subcarrier grouping technique can transmit $G=32$ streams of bit codewords simultaneously through the $32$ groups of two uncorrelated or weakly correlated subcarriers without or with only small performance degradation. 

\ifCLASSOPTIONonecolumn
\begin{figure}[!t]
\centering \includegraphics[width = 1.0\linewidth]{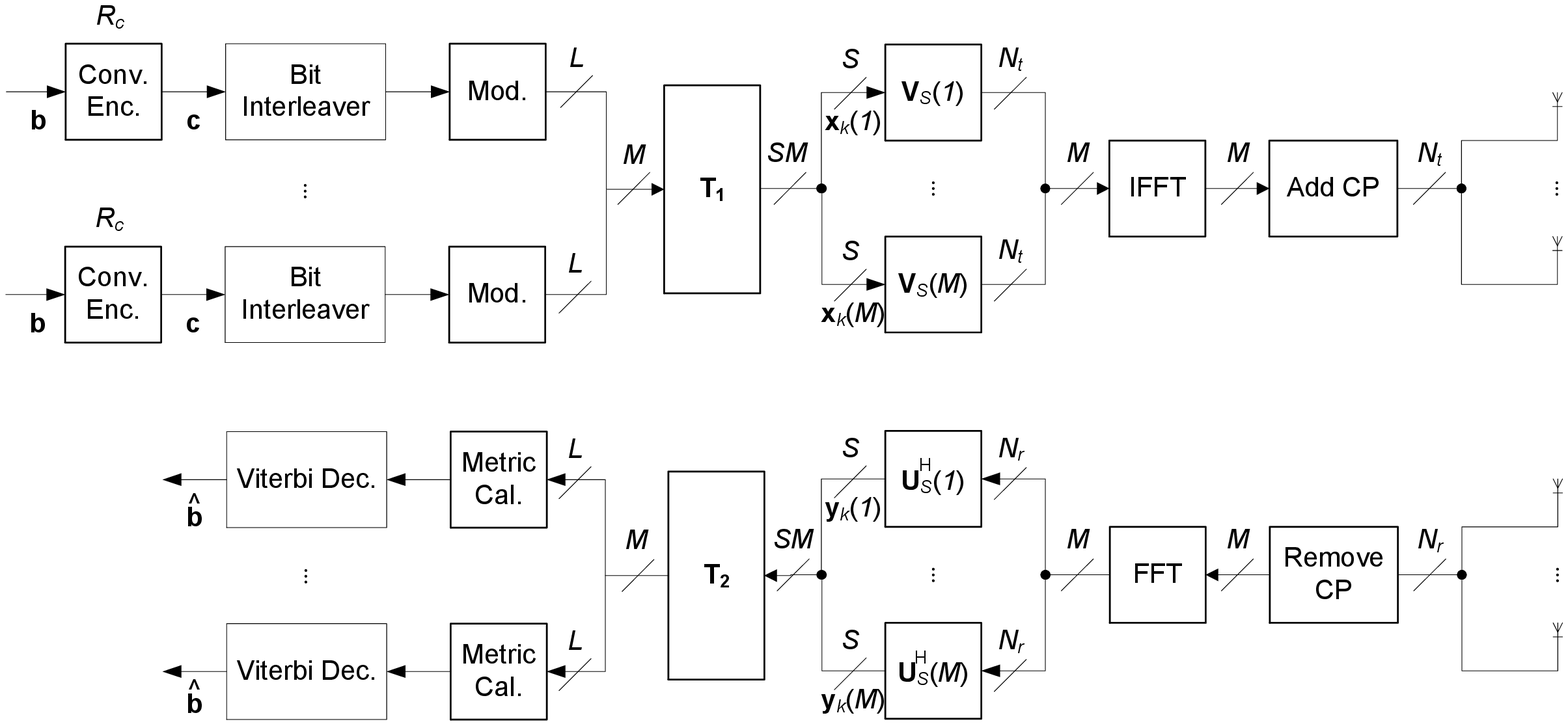}
\caption{Structure of BICMB-OFDM-SG.} \label{fig:bicmb_ofdm_sg}
\end{figure}
\else
\begin{figure}[!t]
\centering \includegraphics[width = 1.0\linewidth]{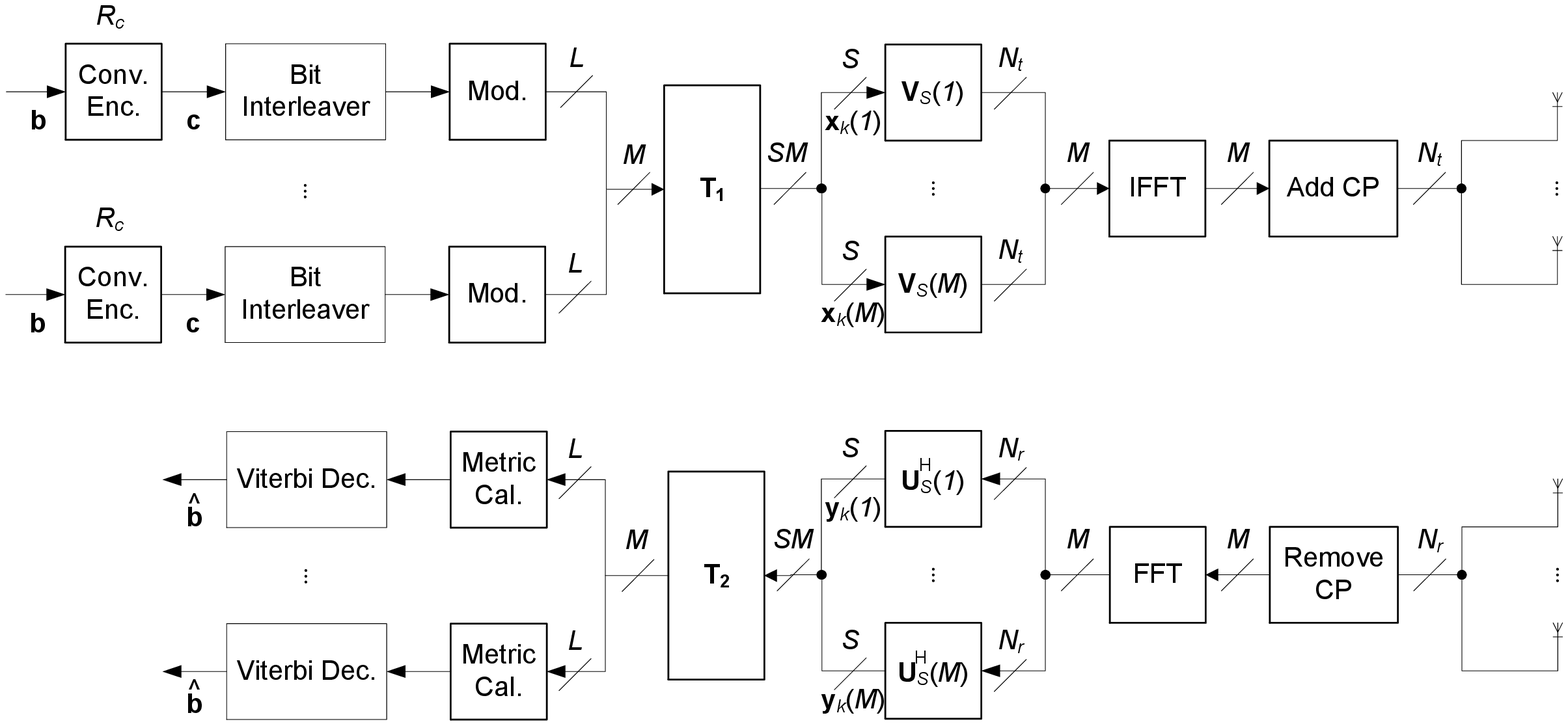}
\caption{Structure of BICMB-OFDM-SG.} \label{fig:bicmb_ofdm_sg}
\end{figure}
\fi

\ifCLASSOPTIONonecolumn
\begin{figure}[!t]
\centering \includegraphics[width = 1.0\linewidth]{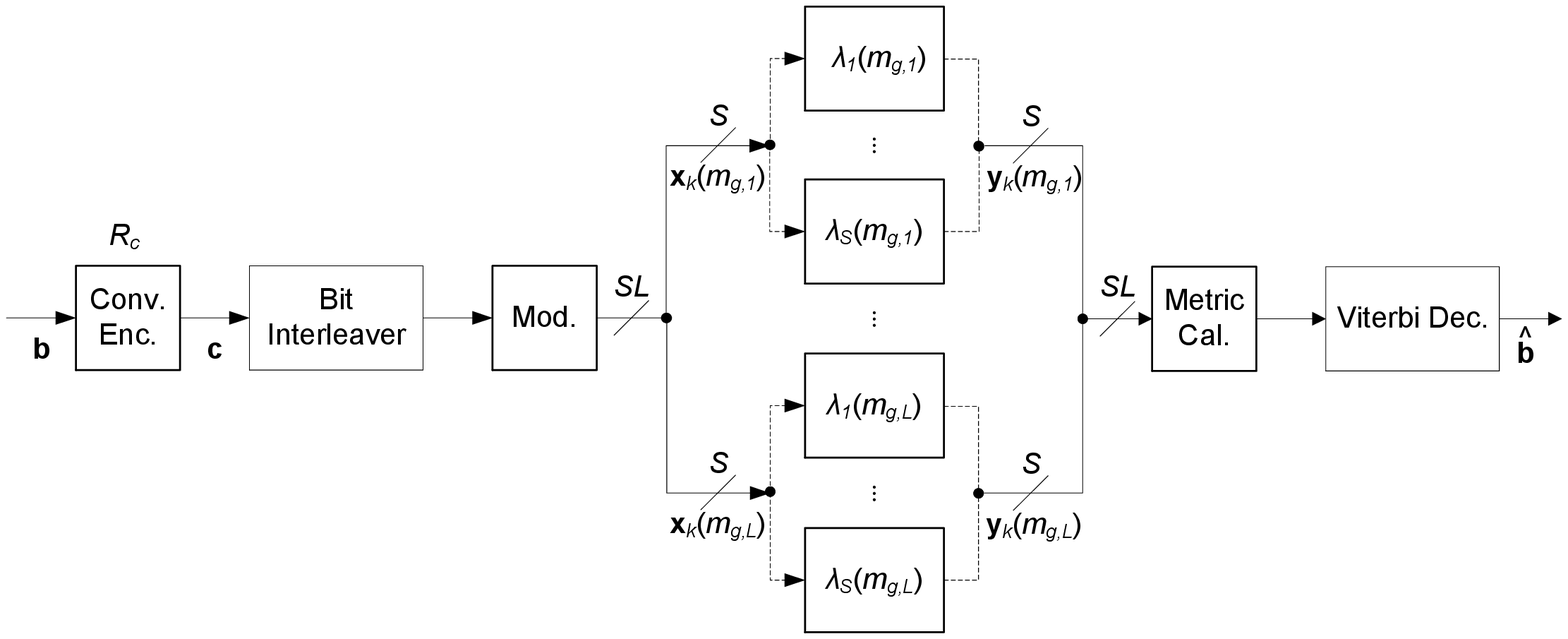}
\caption{Structure of BICMB-OFDM-SG in the frequency domain for one bit stream transmission of the $g$th subcarrier group.} \label{fig:bicmb_ofdm_sgf}
\end{figure}
\else
\begin{figure}[!t]
\centering \includegraphics[width = 1.0\linewidth]{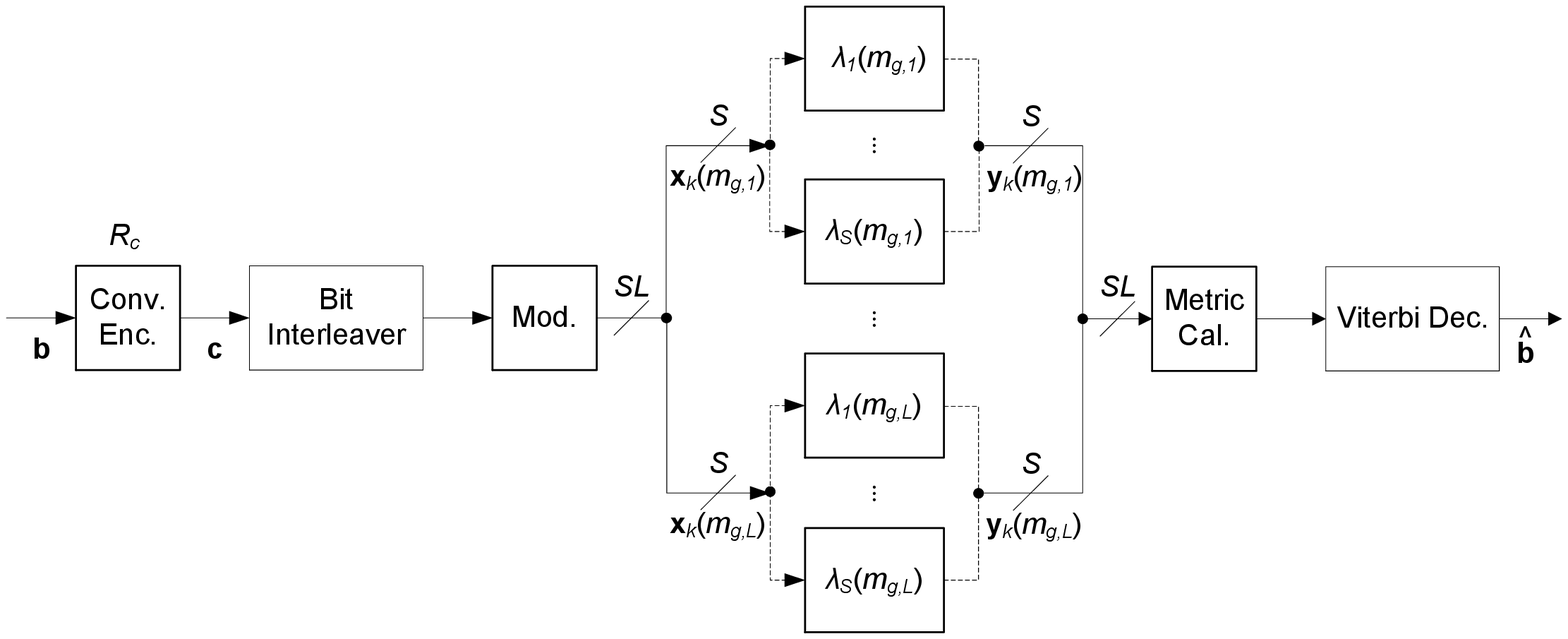}
\caption{Structure of BICMB-OFDM-SG in the frequency domain for one bit stream transmission of the $g$th subcarrier group.} \label{fig:bicmb_ofdm_sgf}
\end{figure}
\fi

Fig. \ref{fig:bicmb_ofdm_sg} presents the structure of BICMB-OFDM-SG. In Fig. \ref{fig:bicmb_ofdm_sg}, $\mathbf{T}_1$ is a permutation matrix at the transmitter that distributes the modulated symbols to their corresponding subcarriers, while $\mathbf{T}_2=\mathbf{T}_1^{-1}$ is a permutation matrix at the receiver that distributes the received symbols to their related streams for decoding. Note that the structure of BICMB-OFDM-SG in Fig. \ref{fig:bicmb_ofdm_sg} can also be considered as Orthogonal Frequency-Division Multiple Access (OFDMA) \cite{Ghosh_LTE} version of BICMB-OFDM. OFDMA is a multi-user version of OFDM and it has been employed in mobile WiMAX \cite{IEEE_802_16} as well as the downlink of LTE \cite{3GPP_TS_36.201}. OFDMA assigns subsets of subcarriers to individual users to achieve multiple access, which is similar to the subcarrier grouping technique. Consequently, BICMB-OFDM can offer multi-user compatibility with subcarrier grouping. Fig. \ref{fig:bicmb_ofdm_sgf} presents the structure of BICMB-OFDM-SG in the frequency domain for one bit stream transmission of the $g$th subcarrier group with $g \in\{1, \ldots, G \}$, and the associated subcarrier index for the $l$th subcarrier of the $g$th group is denoted in the figure as $m_{g,l}=(l-1)G+g$ with $l=1,\ldots,L$. Note that Fig. \ref{fig:bicmb_ofdm_sgf} can also present the structure of BICMB-OFDM in the frequency domain when $L=M$. Compared to BICMB-OFDM without subcarrier grouping, BICMB-OFDM-SG achieves better performance with the same transmission rate and the same decoding complexity while also provides multi-user compatibility. 

\section{System Model of BICMB-OFDM-SG with Precoding} \label{sec:system}

As discussed in Section \ref{subsec:grouping}, BICMB-OFDM-SG is obviously a much better choice than BICMB-OFDM without subcarrier grouping because it provides better performance with the same transmission rate and decoding complexity while also offers multi-user compatibility. Therefore, the precoding technique discussed in the following parts of this paper is employed on top of BICMB-OFDM-SG. 
Since the $G$ groups of bit streams are transmitted separately in the frequency domain for BICMB-OFDM-SG and the only difference is the corresponding singular values of subchannels as shown in Fig. \ref{fig:bicmb_ofdm_sgf}, the precoding technique can be applied to each subcarrier group independently. Therefore, it is sufficient to consider one subcarrier group to illustrate the system model.

\ifCLASSOPTIONonecolumn
\begin{figure}[!t]
\centering \includegraphics[width = 1.0\linewidth]{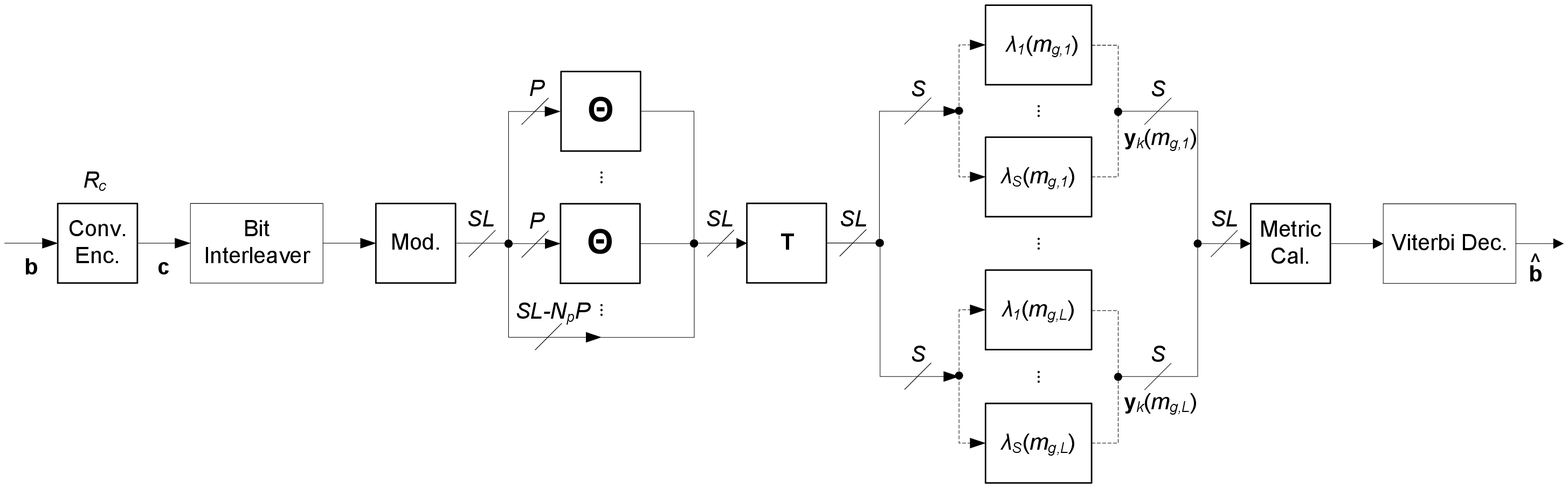}
\caption{Structure of BICMB-OFDM-SG with precoding in the frequency domain for one bit stream transmission of the $g$th subcarrier group.} \label{fig:pbicmb_ofdm}
\end{figure}
\else
\begin{figure}[!t]
\centering \includegraphics[width = 1.0\linewidth]{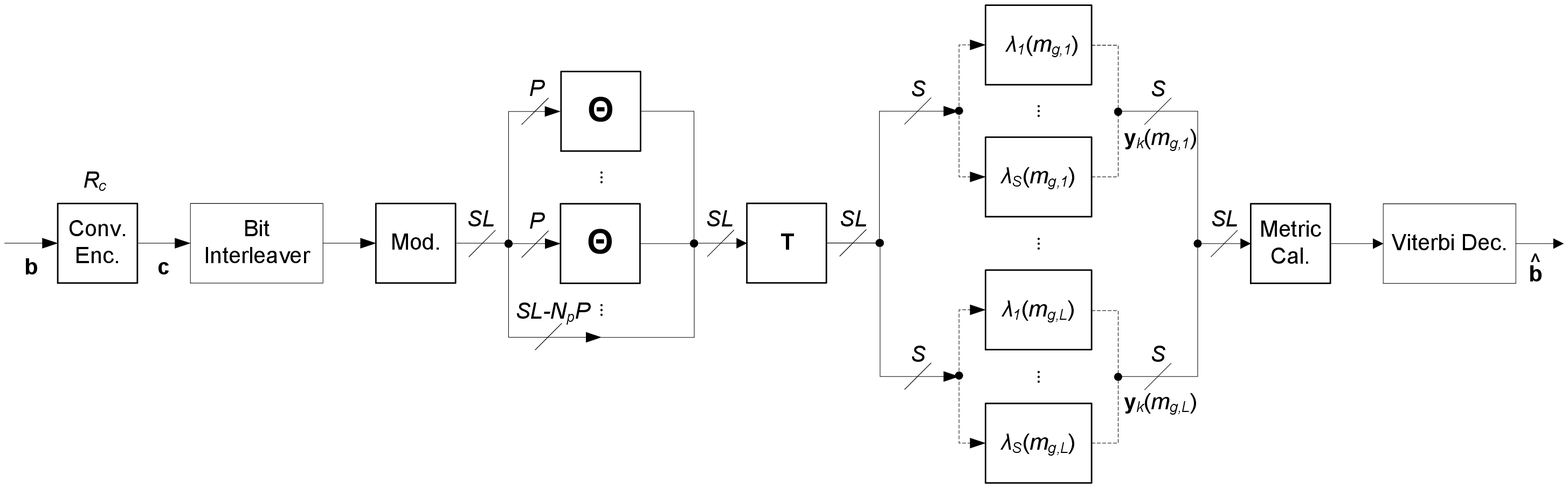}
\caption{Structure of BICMB-OFDM-SG with precoding in the frequency domain for one bit stream transmission of the $g$th subcarrier group.} \label{fig:pbicmb_ofdm}
\end{figure}
\fi

Fig. \ref{fig:pbicmb_ofdm} presents the structure of BICMB-OFDM-SG with precoding in the frequency domain for one bit stream transmission of the $g$th subcarrier group. Compared to BICMB-OFDM-SG without precoding as shown in Fig. \ref{fig:bicmb_ofdm_sgf}, the channel coding, bit interleaver, and modulation remain the same, while two more precoding blocks are added at the transmitter. Specifically, $\mathbf{\Theta}$ is defined as a $P\times P$ precoding matrix where $P\leq SL$ denotes the dimension of $\mathbf{\Theta}$, which is applied to precode $P$ of $SL$ subchannels employed for the $g$th subcarrier group. The $P$ precoded subchannels are defined as one precoded subchannel set. Let $N_p$ denote the number of precoded subchannel sets employed for the $g$th subcarrier group, where $N_pP\leq SL$. As a result, $N_pP$ subchannels are precoded while the remaining $N_n=SL-N_pP$ subchannels are non-precoded. The selections of precoded subchannel sets and non-precoded subchannels are predefined by a permutation matrix $\mathbf{T}$.

Note that there are $L$ subcarriers of each group and each subcarrier includes $S$ subchannels realized by SVD. For the $s$th subchannel at the $l$th subcarrier for the $g$th group, its singular value is $\lambda_s(m_{g,l})$ where $m_{g,l}=(l-1)G+g$. Since the $G$ subcarrier groups are independent, the group index is omitted for brevity in the following, and $\lambda_s(m_{g,l})$ is rewritten as $\lambda_{l,s}$ where the two-dimensional index $\{l,s\}$ denotes the $s$th subchannel at the $l$th subcarrier. For the sake of convenience, the two-dimensional index is further converted to a single dimensional one following the rule $\{l,s\} \rightarrow q=(l-1)S+s$ with $q \in \{1,\ldots,SL\}$, and the corresponding inverse conversion is $q \rightarrow \{l,s\}=\{\left[\lfloor (q-1)/S \rfloor+1\right], \left[(q-1) \bmod S +1\right]\}$. Define $\boldsymbol{\eta}^{z} = \left[\eta_1^{z} \ldots \eta_P^{z} \right]$ as a vector whose elements $\eta_p^{z}$ denote the subchannel indices of the $z$th precoded subchannel set with $z\in\{1,\ldots,N_p\}$, and are ordered increasingly such that $\eta_u^{z} < \eta_v^{z}$ for $u<v$. In the same way, $\boldsymbol{\omega} = \left[ \omega_1 \ldots \omega_{N_n} \right]$ is defined as an increasingly ordered vector whose elements are the indices of the non-precoded subchannels.

\ifCLASSOPTIONonecolumn
At the $k$th time instant, after modulation, the serial-to-parallel converter of the transmitter organizes the $SL \times 1$ symbol vector $\mathbf{x}_k$ as $\mathbf{x}_k = [\mathbf{x}_{\boldsymbol{\eta}^1,k}^T \vdots \ldots \vdots \mathbf{x}_{\boldsymbol{\eta}^{N_p},k}^T \vdots \mathbf{x}_{\boldsymbol{\omega},k}^T]^T $ where $\mathbf{x}_{\boldsymbol{\eta}^z,k} = [x_{\eta_1^z,k} \ldots x_{\eta_P^z,k}]^T$ and $\mathbf{x}_{\boldsymbol{\omega},k} = [x_{\omega_1,k} \ldots x_{\omega_{N_n},k}]^T$ with $x_{q,k}$ denoting the modulated symbol supposed to be transmitted through the $q$th subchannel. Then, at the receiver, the $SL \times 1$ received symbol vector $\mathbf{y}_k = [\mathbf{y}_{\boldsymbol{\eta}^1,k}^T \vdots \ldots \vdots \mathbf{y}_{\boldsymbol{\eta}^{N_p},k}^T \vdots \mathbf{y}_{\boldsymbol{\omega},k}^T]^T $, where $\mathbf{y}_{\boldsymbol{\eta}^z,k} = [y_{\eta_1^z,k} \ldots y_{\eta_P^z,k}]^T$ and $\mathbf{y}_{\boldsymbol{\omega},k} = [y_{\omega_1,k} \ldots y_{\omega_{N_n},k}]^T$ with $y_{q,k}$ denoting the received symbol of the $q$th subchannel, is written as
\begin{align}
\mathbf{y}_k = \breve{\mathbf{\Lambda}} \breve{\mathbf{\Theta}} \mathbf{x}_k + \mathbf{n}_k,
\label{eq:received_vector}
\end{align}
where $\breve{\mathbf{\Lambda}}$ denotes a $SL \times SL$ block diagonal matrix $\breve{\mathbf{\Lambda}} = \textrm{diag}[\breve{\mathbf{\Lambda}}_{\boldsymbol{\eta}^1} \vdots \ldots \vdots \breve{\mathbf{\Lambda}}_{\boldsymbol{\eta}^{N_p}} \vdots \breve{\mathbf{\Lambda}}_{\boldsymbol{\omega}}]$ with diagonal matrices defined as $\breve{\mathbf{\Lambda}}_{\boldsymbol{\eta}^z} = \textrm{diag}[\lambda_{\eta_1^z} \ldots \lambda_{\eta_P^z}]$ and $\breve{\mathbf{\Lambda}}_{\boldsymbol{\omega}} = \textrm{diag}[\lambda_{\omega_1} \ldots \lambda_{\omega_{N_n}}]$, $\breve{\mathbf{\Theta}}$ is a $SL \times SL$ block diagonal matrix $\breve{\mathbf{\Theta}} = \textrm{diag}[\mathbf{\Theta} \vdots \ldots \vdots \mathbf{\Theta} \vdots \mathbf{I}_{N_n}]$ with $\mathbf{I}_{N_n}$ defined as the $N_n$-dimensional identity matrix, and $\mathbf{n}_k$ is an $SL \times 1$ vector $\mathbf{n}_k = [\mathbf{n}_{\boldsymbol{\eta}^1,k}^T \vdots \ldots \vdots \mathbf{n}_{\boldsymbol{\eta}^{N_p},k}^T \vdots \mathbf{n}_{\boldsymbol{\omega},k}^T]^T $, where $\mathbf{n}_{\boldsymbol{\eta}^z,k} = [n_{\eta_1^z,k} \ldots n_{\eta_P^z,k}]^T$ and $\mathbf{n}_{\boldsymbol{\omega},k} = [n_{\omega_1,k} \ldots n_{\omega_{N_n},k}]^T$ with $n_{q,k}$ denoting the additive white Gaussian noise with zero mean and variance $N_0=N_t/\gamma$ at the $q$th subchannel. 
\else
At the $k$th time instant, after modulation, the serial-to-parallel converter of the transmitter organizes the $SL \times 1$ symbol vector $\mathbf{x}_k$ as $\mathbf{x}_k = [\mathbf{x}_{\boldsymbol{\eta}^1,k}^T \vdots \ldots \vdots \mathbf{x}_{\boldsymbol{\eta}^{N_p},k}^T \vdots \mathbf{x}_{\boldsymbol{\omega},k}^T]^T $ where $\mathbf{x}_{\boldsymbol{\eta}^z,k} = [x_{\eta_1^z,k} \ldots x_{\eta_P^z,k}]^T$ and $\mathbf{x}_{\boldsymbol{\omega},k} = [x_{\omega_1,k} \ldots x_{\omega_{N_n},k}]^T$ with $x_{q,k}$ denoting the modulated symbol supposed to be transmitted through the $q$th subchannel. Then, at the receiver, the $SL \times 1$ received symbol vector $\mathbf{y}_k = [\mathbf{y}_{\boldsymbol{\eta}^1,k}^T \vdots \ldots \vdots \mathbf{y}_{\boldsymbol{\eta}^{N_p},k}^T \vdots \mathbf{y}_{\boldsymbol{\omega},k}^T]^T $, where $\mathbf{y}_{\boldsymbol{\eta}^z,k} = [y_{\eta_1^z,k} \ldots y_{\eta_P^z,k}]^T$ and $\mathbf{y}_{\boldsymbol{\omega},k} = [y_{\omega_1,k} \ldots y_{\omega_{N_n},k}]^T$ with $y_{q,k}$ denoting the received symbol of the $q$th subchannel, is written as
\begin{align}
\mathbf{y}_k = \breve{\mathbf{\Lambda}} \breve{\mathbf{\Theta}} \mathbf{x}_k + \mathbf{n}_k,
\label{eq:received_vector}
\end{align}
where $\breve{\mathbf{\Lambda}}$ denotes a $SL \times SL$ block diagonal matrix $\breve{\mathbf{\Lambda}} = \textrm{diag}[\breve{\mathbf{\Lambda}}_{\boldsymbol{\eta}^1} \vdots \ldots \vdots \breve{\mathbf{\Lambda}}_{\boldsymbol{\eta}^{N_p}} \vdots \breve{\mathbf{\Lambda}}_{\boldsymbol{\omega}}]$ with diagonal singular matrices defined as $\breve{\mathbf{\Lambda}}_{\boldsymbol{\eta}^z} = \textrm{diag}[\lambda_{\eta_1^z} \ldots \lambda_{\eta_P^z}]$ and $\breve{\mathbf{\Lambda}}_{\boldsymbol{\omega}} = \textrm{diag}[\lambda_{\omega_1} \ldots \lambda_{\omega_{N_n}}]$, $\breve{\mathbf{\Theta}}$ is a $SL \times SL$ block diagonal matrix $\breve{\mathbf{\Theta}} = \textrm{diag}[\mathbf{\Theta} \vdots \ldots \vdots \mathbf{\Theta} \vdots \mathbf{I}_{N_n}]$ with $\mathbf{I}_{N_n}$ defined as the $N_n$-dimensional identity matrix, and $\mathbf{n}_k$ is an $SL \times 1$ vector $\mathbf{n}_k = [\mathbf{n}_{\boldsymbol{\eta}^1,k}^T \vdots \ldots \vdots \mathbf{n}_{\boldsymbol{\eta}^{N_p},k}^T \vdots \mathbf{n}_{\boldsymbol{\omega},k}^T]^T $, where $\mathbf{n}_{\boldsymbol{\eta}^z,k} = [n_{\eta_1^z,k} \ldots n_{\eta_P^z,k}]^T$ and $\mathbf{n}_{\boldsymbol{\omega},k} = [n_{\omega_1,k} \ldots n_{\omega_{N_n},k}]^T$ with $n_{q,k}$ denoting the additive white Gaussian noise with zero mean and variance $N_0=N_t/\gamma$ at the $q$th subchannel.
\fi
As a result, the input-output relation in (\ref{eq:received_vector}) can be decomposed into $N_p+1$ equations as
\begin{align}
\begin{split}
& \mathbf{y}_{\boldsymbol{\eta}^z,k} = \breve{\mathbf{\Lambda}}_{\boldsymbol{\eta}^z} \mathbf{\Theta} \mathbf{x}_{\boldsymbol{\eta}^z,k} + \mathbf{n}_{\boldsymbol{\eta}^z,k}, \\
& \mathbf{y}_{\boldsymbol{\omega},k} = \breve{\mathbf{\Lambda}}_{\boldsymbol{\omega}} \mathbf{x}_{\boldsymbol{\omega},k} + \mathbf{n}_{\boldsymbol{\omega},k}.
\label{eq:received_decomposed}
\end{split}
\end{align}

In a similar way to BICMB-OFDM introduced in Section \ref{sec:basis}, the location of the coded bit $c_{k'}$ within the transmitted symbol is denoted as $k' \rightarrow (k, q, j)$, which means that $c_{k'}$ is mapped onto the $j$th bit position on the label of $x_{q,k}$. By using the location information and the input-output relation in (\ref{eq:received_vector}), the receiver calculates the ML bit metrics for $c_{k'}=b \in \{0, 1\}$ as
\begin{align}
\Delta(\mathbf{y}_k, c_{k'}) = \min_{\mathbf{x} \in \xi_{c_{k'}}^{q,j}} \| \mathbf{y}_k - \breve{\mathbf{\Lambda}} \breve{\mathbf{\Theta}} \mathbf{x} \|^2, \label{eq:ml_bit_metrics}
\end{align}
where $\xi_{c_{k'}}^{q,j}$ is a subset of $\chi^{SL}$, defined as
\begin{align*}
\xi_{b}^{q,j} = \{ \mathbf{x} = [x_1 \ldots x_{SL} ]^T: x_{u=q} \in \chi_{b}^{j} \, \textrm{and} \, x_{u \neq q} \in \chi \}.
\end{align*}
Based on the decomposition of (\ref{eq:received_vector}) to (\ref{eq:received_decomposed}), the bit metrics equivalent
to (\ref{eq:ml_bit_metrics}) are
\begin{align}
\Delta(\mathbf{y}_k, c_{k'}) =  \left\{
\begin{array}{ll}
\min\limits_{\mathbf{x} \in \psi_{c_{k'}}^{\hat{q},j}} \| \mathbf{y}_{\boldsymbol{\eta}^z,k} - \breve{\mathbf{\Lambda}}_{\eta^z} \mathbf{\Theta} \mathbf{x} \|^2, & \textrm{if $q \in \boldsymbol{\eta}^z$}, \\
\min\limits_{x \in \chi_{c_{k'}}^{j}} \left| y_{q,k} - \lambda_q x \right|^2, & \textrm{if $q \in \boldsymbol{\omega}$},
\end{array} \right.
\label{eq:ml_bit_metrics_decomposed}
\end{align}
where $\hat{q}\in\{1,\ldots,P\}$ is the associated index of the $q$th subchannel in $\boldsymbol{\eta}^z$, and $\psi_{b}^{\hat{q},j}$ is a subset of $\chi^P$ defined as
\begin{equation*}
\psi_{b}^{\hat{q},j} = \{ \mathbf{x} = [x_1 \ldots x_P ]^T : x_{p=\hat{q}} \in \chi_{b}^{j} \, \textrm{and} \, x_{p \neq \hat{q}} \in \chi
\}.
\end{equation*}

Finally, the ML decoder, which applies the soft-input Viterbi decoding to find a codeword $\mathbf{\hat{c}}$ with the minimum sum weight and its corresponding information bit sequence $\mathbf{\hat{b}}$, uses the bit metrics calculated by (\ref{eq:ml_bit_metrics_decomposed}) and makes decisions as
\begin{align}
\mathbf{\hat{c}} = \arg\min_{\mathbf{c}} \sum_{k'} \Delta(\mathbf{y}_k, c_{k'}).
\label{eq:decision_rule}
\end{align}

O\section{Diversity Analysis of BICMB-OFDM-SG with Precoding} \label{sec:diversity} 

The performance of BICMB-OFDM-SG with precoding of each subcarrier group is bounded by the union of the Pairwise Error Probability (PEP) corresponding to each error event \cite{Akay_BICMB_OFDM, Akay_BICMB_OFDM_CSI, Akay_BICMB}. In particular, the overall diversity order is dominated by the pairwise errors which have the smallest negative exponent value of SNR in their PEP representations. As a result, the calculation of each PEP is needed. In this section, an upper bound to each PEP is derived.

Based on the bit metrics in (\ref{eq:ml_bit_metrics}), the instantaneous PEP between the transmitted codeword $\mathbf{c}$ and
the decoded codeword $\mathbf{\hat{c}}$ is calculated as
\begin{align}
\mathrm{Pr} \left( \mathbf{c} \rightarrow \hat{\mathbf{c}} \mid \mathbf{H}(m), \forall m \right) = \mathrm{Pr} \left( \sum_{k'} \min_{\mathbf{x} \in \xi_{c_{k'}}^{q,j}} \| \mathbf{y}_k - \breve{\mathbf{\Lambda}} \breve{\mathbf{\Theta}} \mathbf{x} \|^2 \geq  \ifCLASSOPTIONtwocolumn \right. \nonumber \\ \left. \fi \sum_{k'} \min_{\mathbf{x} \in \xi_{\hat{c}_{k'}}^{q,j}} \| \mathbf{y}_k - \breve{\mathbf{\Lambda}} \breve{\mathbf{\Theta}} \mathbf{x} \|^2 \right),
\label{eq:pep_original}
\end{align}
where $c_{k'}$ and $\hat{c}_{k'}$ are the coded bits of $\mathbf{c}$ and $\mathbf{\hat{c}}$, respectively. Let $d_H$ denote the Hamming distance \cite{Lin_ECC} between $\mathbf{c}$ and $\mathbf{\hat{c}}$. Since the bit metrics corresponding to the same coded bits between the pairwise errors are the same, (\ref{eq:pep_original}) is rewritten as
\begin{align}
\mathrm{Pr} \left(\mathbf{c} \rightarrow \hat{\mathbf{c}} \mid \mathbf{H}(m), \forall m \right) = \mathrm{Pr} \left( \sum_{k', d_H} \min_{\mathbf{x} \in \xi_{c_{k'}}^{q,j}} \| \mathbf{y}_k - \breve{\mathbf{\Lambda}} \breve{\mathbf{\Theta}} \mathbf{x} \|^2 \geq \right. \ifCLASSOPTIONtwocolumn \nonumber \\ \fi \left. \sum_{k', d_H} \min_{\mathbf{x} \in \xi_{\hat{c}_{k'}}^{q,j}} \| \mathbf{y}_k - \breve{\mathbf{\Lambda}} \breve{\mathbf{\Theta}} \mathbf{x} \|^2 \right),
\label{eq:pep_diffbits}
\end{align}
where $\sum_{k', d_H}$ stands for the summation of the $d_H$ values corresponding to the different coded bits between the bit codewords.

Define $\tilde{\mathbf{x}}_k$ and $\hat{\mathbf{x}}_k$ as
\begin{align}
\begin{split}
\tilde{\mathbf{x}}_k  = \arg \limits{\min}_{\mathbf{x} \in \xi_{c_{k'}}^{q,j}} \| \mathbf{y}_k - \breve{\mathbf{\Lambda}} \breve{\mathbf{\Theta}} \mathbf{x} \|^2, \\
\hat{\mathbf{x}}_k  = \arg\limits{\min}_{\mathbf{x} \in \xi_{\bar{c}_{k'}}^{q,j}} \| \mathbf{y}_k - \breve{\mathbf{\Lambda}} \breve{\mathbf{\Theta}} \mathbf{x} \|^2,
\end{split}
\label{eq:arg_min}
\end{align}
where $\bar{c}_{k'}$ is the complement of $c_{k'}$ in binary. It is easily found that $\tilde{\mathbf{x}}_k$ is different from $\hat{\mathbf{x}}_k$ since the sets that $x_q$ belong to are disjoint, as can be seen from the definition of $\xi_{c_{k'}}^{q,j}$. In the same manner, it is clear that $\mathbf{x}_k$ is different from $\hat{\mathbf{x}}_k$. With $\tilde{\mathbf{x}}_k$ and $\hat{\mathbf{x}}_k$, (\ref{eq:pep_diffbits}) is rewritten as
\begin{align}
\mathrm{Pr} \left( \mathbf{c} \rightarrow \mathbf{\hat{c}} \mid \mathbf{H}(m), \forall m \right) = \mathrm{Pr} \left( \sum_{k', d_H} \| \mathbf{y}_k - \breve{\mathbf{\Lambda}} \breve{\mathbf{\Theta}} \tilde{\mathbf{x}}_k \|^2 \geq \right. \ifCLASSOPTIONtwocolumn \nonumber \\ \fi \left. \sum_{k', d_H} \| \mathbf{y}_k - \breve{\mathbf{\Lambda}} \breve{\mathbf{\Theta}} \hat{\mathbf{x}}_k \|^2 \right).
\label{eq:alt_pep_diffbits}
\end{align}
Based on the fact that $\| \mathbf{y}_k - \breve{\mathbf{\Lambda}} \breve{\mathbf{\Theta}} \mathbf{x}_k \|^2 \geq  \| \mathbf{y}_k - \breve{\mathbf{\Lambda}} \breve{\mathbf{\Theta}} \tilde{\mathbf{x}}_k \|^2$, and the relation in (\ref{eq:received_vector}), equation (\ref{eq:alt_pep_diffbits}) is upper bounded by
\ifCLASSOPTIONonecolumn
\begin{align}
\mathrm{Pr} \left(\mathbf{c} \rightarrow \mathbf{\hat{c}} \mid \mathbf{H}(m), \forall m \right) & \leq \mathrm{Pr} \left( \sum_{k', d_H} \| \mathbf{y}_k - \breve{\mathbf{\Lambda}} \breve{\mathbf{\Theta}} {\mathbf{x}}_k \|^2 \geq \sum_{k', d_H} \| \mathbf{y}_k - \breve{\mathbf{\Lambda}} \breve{\mathbf{\Theta}} \hat{\mathbf{x}}_k \|^2 \right) \nonumber \\ 
& = \mathrm{Pr} \left( \epsilon \geq \sum_{k', d_H} \| \breve{\mathbf{\Lambda}} \breve{\mathbf{\Theta}} (\mathbf{x}_k - \mathbf{\hat{x}}_k) \|^2 \right), \label{eq:pep_upper}
\end{align}
\else
\begin{align}
\mathrm{Pr} \left(\mathbf{c} \rightarrow \mathbf{\hat{c}} \mid \mathbf{H}(m), \forall m \right) & \leq \mathrm{Pr} \left( \sum_{k', d_H} \| \mathbf{y}_k - \breve{\mathbf{\Lambda}} \breve{\mathbf{\Theta}} {\mathbf{x}}_k \|^2 \geq \right. \nonumber \\ 
& \qquad \quad \left. \sum_{k', d_H} \| \mathbf{y}_k - \breve{\mathbf{\Lambda}} \breve{\mathbf{\Theta}} \hat{\mathbf{x}}_k \|^2 \right) \nonumber \\ 
& = \mathrm{Pr} \left( \epsilon \geq \sum_{k', d_H} \| \breve{\mathbf{\Lambda}} \breve{\mathbf{\Theta}} (\mathbf{x}_k - \mathbf{\hat{x}}_k) \|^2 \right), \label{eq:pep_upper}
\end{align}
\fi
where $\epsilon = \sum_{k', d_H} \mathrm{Tr} [ - (\mathbf{x}_k - \hat{\mathbf{x}_k})^H \breve{\mathbf{\Theta}}^H \breve{\mathbf{\Lambda}}^H \mathbf{n}_k - \mathbf{n}_k^H \breve{\mathbf{\Lambda}} \breve{\mathbf{\Theta}} (\mathbf{x}_k - \hat{\mathbf{x}}_k) ]$. Since $\epsilon$ is a zero-mean Gaussian random variable with variance $2 N_0 \sum_{k', d_H} \| \breve{\mathbf{\Lambda}} \breve{\mathbf{\Theta}}  (\mathbf{x}_k - \mathbf{\hat{x}}_k) \| ^2$, (\ref{eq:pep_upper}) is given by the $\mathrm{Q}$ function as
\begin{align}
\mathrm{Pr} \left(\mathbf{c} \rightarrow \mathbf{\hat{c}} \mid \mathbf{H}(m), \forall m \right) \leq \mathrm{Q} \left( \sqrt{\frac{\| \breve{\mathbf{\Lambda}} \breve{\mathbf{\Theta}} (\mathbf{x}_k - \mathbf{\hat{x}}_k) \| ^2}{2 N_0}}\right). \label{eq:pep_upper_q}
\end{align}
By using the upper bound on the $\mathrm{Q}$ function $\mathrm{Q}(x) \leq 0.5 \exp(-x^2/2)$, the average PEP can be upper bounded as
\begin{align}
\mathrm{Pr} \left( \mathbf{c} \rightarrow \mathbf{\hat{c}} \right) & = \mathrm{E} \left[ \mathrm{Pr} \left( \mathbf{c} \rightarrow \mathbf{\hat{c}} \mid \mathbf{H}(m), \forall m \right) \right] \nonumber \\
& \leq \mathrm{E} \left[ \frac{1}{2} \exp \left(- \frac{\sum_{k', d_H} \| \breve{\mathbf{\Lambda}} \breve{\mathbf{\Theta}} (\mathbf{x}_k - \mathbf{\hat{x}}_k) \| ^2}{4 N_0} \right) \right]. 
\label{eq:pep_average}
\end{align}

According to (\ref{eq:received_decomposed}), the negative numerator of the exponent in (\ref{eq:pep_average}) is rewritten as
\ifCLASSOPTIONtwocolumn
\begin{align}
\kappa & = \sum_{k', d_H} \| \breve{\mathbf{\Lambda}} \breve{\mathbf{\Theta}} (\mathbf{x}_k - \mathbf{\hat{x}}_k) \| ^2 \nonumber \\
& = \sum_{z=1}^{N_p} \sum_{k', d_{H,\boldsymbol{\eta}^z}} \| \breve{\mathbf{\Lambda}}_{\boldsymbol{\eta}^z} \mathbf{\Theta} \left( \mathbf{x}_{\boldsymbol{\eta}^z,k} - \hat{\mathbf{x}}_{\boldsymbol{\eta}^z,k} \right) \|^2 + \nonumber \\ 
& \quad \sum_{u=1}^{N_n} \sum_{k', d_{H,\omega_u}} | \lambda_{\omega_u} \left( x_{\omega_u,k} - \hat{x}_{\omega_u,k} \right) |^2 \nonumber \\
& = \sum_{z=1}^{N_p} \sum_{p=1}^{P} \lambda_{\eta^z_p}^2 \sum_{k', d_{H,\boldsymbol{\eta}^z}} | \boldsymbol{\theta}_p^T \left( \mathbf{x}_{\boldsymbol{\eta}^z,k} - \hat{\mathbf{x}}_{\boldsymbol{\eta}^z,k} \right) |^2 + \nonumber \\ 
& \quad \sum_{u=1}^{N_n} \lambda_{\omega_u}^2 \sum_{k', d_{H,\omega_u}} | \left( x_{\omega_u,k} - \hat{x}_{\omega_u,k} \right) |^2,
\label{eq:pep_numerator}
\end{align}
\else
\begin{align}
\kappa & = \sum_{k', d_H} \| \breve{\mathbf{\Lambda}} \breve{\mathbf{\Theta}} (\mathbf{x}_k - \mathbf{\hat{x}}_k) \| ^2 \nonumber \\
& = \sum_{z=1}^{N_p} \sum_{k', d_{H,\boldsymbol{\eta}^z}} \| \breve{\mathbf{\Lambda}}_{\boldsymbol{\eta}^z} \mathbf{\Theta} \left( \mathbf{x}_{\boldsymbol{\eta}^z,k} - \hat{\mathbf{x}}_{\boldsymbol{\eta}^z,k} \right) \|^2 + \sum_{u=1}^{N_n} \sum_{k', d_{H,\omega_u}} | \lambda_{\omega_u} \left( x_{\omega_u,k} - \hat{x}_{\omega_u,k} \right) |^2 \nonumber \\
& = \sum_{z=1}^{N_p} \sum_{p=1}^{P} \lambda_{\eta^z_p}^2 \sum_{k', d_{H,\boldsymbol{\eta}^z}} | \boldsymbol{\theta}_p^T \left( \mathbf{x}_{\boldsymbol{\eta}^z,k} - \hat{\mathbf{x}}_{\boldsymbol{\eta}^z,k} \right) |^2 + \sum_{u=1}^{N_n} \lambda_{\omega_u}^2 \sum_{k', d_{H,\omega_u}} | \left( x_{\omega_u,k} - \hat{x}_{\omega_u,k} \right) |^2,
\label{eq:pep_numerator}
\end{align}
\fi
where $\sum_{k', d_{H,\boldsymbol{\eta}^z}}$ and $\sum_{k', d_{H,\omega_u}}$ stand for the summation related to the $d_{H,\boldsymbol{\eta}^z}$ and $d_{H,\omega_u}$ different coded bits carried on the subchannels in $\boldsymbol{\eta}^z$ and subchannel $\omega_u$, respectively, and $\boldsymbol{\theta}_p^T$ denotes the $p$th row of $\mathbf{\Theta}$. By reordering the indices of singular values, (\ref{eq:pep_numerator}) can be rewritten as the following form
\begin{align}
\kappa = \sum_{q=1}^{SL} \rho_q \lambda_q^2,
\label{eq:pep_numerator_ordered}
\end{align}
where 
\begin{align}
\rho_q =  \left\{
\begin{array}{ll}
\sum\limits_{k', d_{H,\boldsymbol{\eta}^z}} | \boldsymbol{\theta}_{\hat{q}}^T \left( \mathbf{x}_{\boldsymbol{\eta}^z,k} - \hat{\mathbf{x}}_{\boldsymbol{\eta}^z,k} \right) |^2, & \textrm{if $q \in \boldsymbol{\eta}^z$}, \\
\sum\limits_{k', d_{H,\omega_u}} | \left( x_{\omega_u,k} - \hat{x}_{\omega_u,k} \right) |^2, & \textrm{if $q = \omega_u$},
\end{array} \right.
\label{eq:weight_lambda}
\end{align}
with $\hat{q}$ denoting the associated index of the $q$th subchannel in its precoded subchannel set. For BICMB-OFDM-SG, the subcarriers of each group are uncorrelated or weakly correlated \cite{Li_DA_BICMB_OFDM}. In that case, the singular value matrices $\mathbf{\Lambda}(m)$ can be considered independent for each subcarrier group \cite{Li_DA_BICMB_OFDM}. Therefore, by converting the one-dimensional subchannel indices back to their corresponding two-dimensional indices, (\ref{eq:pep_average}) is further rewritten as 
\begin{align}
\mathrm{Pr} \left( \mathbf{c} \rightarrow \mathbf{\hat{c}} \right) & \leq \prod_{l} \mathrm{E} \left[ \exp \left(- \frac{ \sum_{s} \rho_{l,s} {\lambda}_{l,s}^2} {4 N_0} \right) \right].
\label{eq:pep_uncorrelated}
\end{align}
For each subcarrier, the terms inside the expectation in (\ref{eq:pep_uncorrelated}) can be upper bounded by employing the theorem proved in \cite{Park_UP_MPDF}, which is given below.
\begin{theorem}
Consider the largest $S \leq \min\{N_t, N_r\}$ eigenvalues $\mu_s$ of the uncorrelated central $N_r \times N_t$ Wishart matrix \cite{Edelman_Eigen} that are sorted in decreasing order, and a weight vector $\boldsymbol{\rho} = [\rho_1 \ldots \rho_S]^T$ with non-negative real elements. In the high SNR regime, an upper bound for the expression $\mathrm{E} [ \exp (-\gamma \sum_{s=1}^S \rho_s \mu_s ) ]$, which is used in the diversity analysis of a number of MIMO systems, is
\begin{align}
\mathrm{E} \left[ \exp \left( - \gamma \sum\limits_{s=1}^S \rho_s \mu_s \right) \right] \leq \zeta \left( \rho_{min} \gamma \right)^{-(N_r-\delta+1)(N_t-\delta+1)},
\label{eq:theorem_e_pep}
\end{align}
where $\gamma$ is SNR, $\zeta$ is a constant, $\rho_{min} = \min_{\rho_i \neq 0} {\{ \rho_i \}}_{i=1}^{S}$, and $\delta$ is the index to the first non-zero element in the weight vector.
\label{theorem:e_pep}
\end{theorem}
\begin{IEEEproof}
See \cite{Park_UP_MPDF}.
\end{IEEEproof}
By applying the aforementioned theorem to (\ref{eq:pep_uncorrelated}), an upper bound of PEP is
\begin{align}
\mathrm{Pr} \left( \mathbf{c} \rightarrow \mathbf{\hat{c}} \right) & \leq \prod_{l,\boldsymbol{\rho}_l \neq \mathbf{0}} {\zeta}_l \left( \frac{ \rho_{l,min}} {4 N_t} \gamma \right)^{-D_l},
\label{eq:pep_diversity}
\end{align}
with $D_l=(N_r-\delta_l+1)(N_t-\delta_l+1)$, where $\rho_{l,min}$ denotes the minimum non-zero element in $\boldsymbol{\rho}_l$ whose element $\rho_{l,s}$ denotes the weight of $\lambda^2_{l,s}$, $\delta_l$ denotes the index of the first non-zero element in $\boldsymbol{\rho}_l$, and ${\zeta}_l$ is a constant. Therefore, the diversity can be easily found from (\ref{eq:pep_diversity}), which is 
\begin{align}
D = \sum_{l,\boldsymbol{\rho}_l \neq \mathbf{0}} D_l.
\label{eq:diversity}
\end{align}
Based on the results of (\ref{eq:pep_diversity}) and (\ref{eq:diversity}), full diversity 
can be achieved if and only if $\rho_{l,1} \neq 0, \, \forall l$ for all error events. Since the error events are related to the convolutional code and the bit interleaver, the full diversity condition is related to the combination of the precoding matrix, the convolutional code, and the bit interleaver.

\section{Full-Diversity Precoding Design of BICMB-OFDM-SG with Precoding} \label{sec:design} 

The precoding design satisfying the full diversity condition $\rho_{l,1} \neq 0, \, \forall l$ of all error events may not be unique.
In this section, a sufficient method of precoding design is developed for BICMB-OFDM-SG which guarantees full diversity while minimizing the increased decoding complexity caused by precoding. 

\subsection{Choice of Precoding Matrix} \label{subsec:pre_type}
An upper bound of PEP for BICMB-OFDM-SG without precoding can be written as in \cite{Li_DA_BICMB_OFDM} in a similar form as (\ref{eq:pep_uncorrelated}) only with different weights of
\begin{align}
\tilde{\rho}_{l,s} =  d^2_{min} \alpha_{l,s},
\label{eq:weight_lambda_alpha}
\end{align}
where $d_{min}$ is the minimum Euclidean distance \cite{Barry_DC} in the constellation, and $\alpha_{l,s}$ denotes the number of distinct bits transmitting through the $s$th subchannel of the $l$th subcarrier for an error path which implies that $\sum_{l=1}^L \sum_{s=1}^S \alpha_{l,s} = d_H$. The diversity can be derived in a similar fashion to (\ref{eq:pep_diversity}) and (\ref{eq:diversity}), and the full diversity condition is $\alpha_{l,1} \neq 0, \, \forall l$ for all error events. As proved in \cite{Li_DA_BICMB_OFDM}, the full diversity condition can be achieved only if the condition of $R_cSL \leq 1$ is satisfied. Otherwise, full diversity cannot be provided. The reason is that, in the case of $R_cSL > 1$, there always exists at least one error path with no errored bit of the error event transmitted through the first subchannel of a subcarrier. 

It is obvious that when $\alpha_{l,1} = 0$, then $\rho_{l,1} = 0$, if the $\{l,1\}$th subchannel is non-precoded. However, if the $\{l,1\}$th subchannel is precoded, $\rho_{l,1}$ could be non-zero even if $\alpha_{l,1} = 0$, which depends on $\boldsymbol{\theta}_{\hat{q}}^T$ and each error event as shown in (\ref{eq:weight_lambda}). Therefore, BICMB-OFDM-SG with precoding could achieve full diversity even if $R_cSL > 1$ by proper precoding design. When designing the precoding matrix, it is inconvenient to consider all error events which could be large in number. However, since an error event only affects $ \mathbf{x}_{\boldsymbol{\eta}^z,k} - \hat{\mathbf{x}}_{\boldsymbol{\eta}^z,k}$ in (\ref{eq:weight_lambda}), a sufficient condition of the precoding design is given by
\begin{align}
| \boldsymbol{\theta}_{\hat{q}}^T \left( \mathbf{x} - \hat{\mathbf{x}} \right) |^2 \neq 0, & \, \, \textrm{for $\left(q \bmod S \right) = 1$},
\label{eq:pre_design_prerow}
\end{align}
of all different $\mathbf{x}$ and $\hat{\mathbf{x}}$. It is not hard to find $\boldsymbol{\theta}_{\hat{q}}^T$ which satisfies (\ref{eq:pre_design_prerow}). In fact, as long as every element in $\boldsymbol{\theta}_{\hat{q}}^T$ is non-zero, the condition (\ref{eq:pre_design_prerow}) is satisfied. 

Note that the condition (\ref{eq:pre_design_prerow}) is designed for certain rows of $\mathbf{\Theta}$ corresponding to the first subchannel of all subcarriers. Although other subchannels do not affect the diversity as shown in (\ref{eq:pep_diversity}) and (\ref{eq:diversity}), the condition (\ref{eq:pre_design_prerow}) can be further simplified to 
\begin{align}
| \boldsymbol{\theta}_{p}^T \left( \mathbf{x} - \hat{\mathbf{x}} \right) |^2 \neq 0, \forall p,
\label{eq:pre_design_row}
\end{align}
of all different $\mathbf{x}$ and $\hat{\mathbf{x}}$. 

Assume that the average transmitted power at each transmit antenna is the same, then the precoding matrix is chosen as
\begin{align}
\theta_{u,v} \neq 0, \forall u, \forall v & \, \, \textrm{and $\| \boldsymbol{\theta}_{p}^T \|^2 = 1, \forall p$}.
\label{eq:pre_design}
\end{align}
In fact, the precoding matrices in \cite{Park_CPB} all satisfy the condition (\ref{eq:pre_design}), which are considered in the next three subsections.

\subsection{Minimum Effective Dimension of Precoding Matrix} \label{subsec:pre_dim}

When the precoding matrices in \cite{Park_CPB} are applied, the weights of (\ref{eq:weight_lambda}) can be simplified to
\begin{align}
\rho_{q} =  d_{min}^2 \beta_q, 
\label{eq:weight_lambda_beta}
\end{align}
where
\begin{align}
\beta_{q} =  \left\{
\begin{array}{ll}
|\theta_{\hat{q},min}|^2 \alpha_{\boldsymbol{\eta}^z,min}, & \textrm{if $q \in \boldsymbol{\eta}_z$}, \\
\alpha_q, & \textrm{if $q = \omega_u$},
\end{array} \right.
\label{eq:beta}
\end{align}
with $\theta_{\hat{q},min}$ denoting the element in $\boldsymbol{\theta}_{\hat{q}}$ having the smallest absolute value and $\alpha_{\boldsymbol{\eta}^z,min}$ denoting the minimum non-zero $\alpha$ element corresponding to the $P$ precoded subchannels for the $z$th set. Note that (\ref{eq:weight_lambda_beta}) is a lower bound for (\ref{eq:weight_lambda}). As a result, an upper bound with a form similar to (\ref{eq:pep_diversity}) and (\ref{eq:diversity}) can be derived with the simplified weights (\ref{eq:weight_lambda_beta}). Compared to the weights of (\ref{eq:weight_lambda_alpha}) for BICMB-OFDM-SG without precoding, the weights of the $N_n$ non-precoded subchannels are the same. For the $N_pP$ precoded subchannels, each weight now depends on the $\alpha$ elements of the $P$ precoded subchannels of the corresponding set instead of only one subchannel. Therefore, if an errored bit is transmitted through a precoded subchannel, then all weights in (\ref{eq:weight_lambda_beta}) for the $P$ precoded subchannels of the corresponding set is non-zero. However, if no errored bit is transmitted through a precoded subchannel set, then all weights of the $P$ precoded subchannels are zero, which are the same as BICMB-OFDM-SG without precoding. If that happens, precoding is meaningless since the PEPs with the worst diversity dominate the overall performance. Therefore, the precoding design requirement is that at least an errored bit of each error event is transmitted through each precoded subchannel set.

The aforementioned precoding design requirement is related to the convolutional code, the bit interleaver, and the dimension of the precoding matrix. In fact, if $P=SL$, which means all subchannels are precoded by only one $SL \times SL$ precoding matrix $\mathbf{\Theta}$, the requirement can be easily satisfied. However, a larger dimension for $\mathbf{\Theta}$ results in higher complexity for calculating the metrics associated with the precoded bits in (\ref{eq:ml_bit_metrics_decomposed}). As a result, the minimum effective dimension of $\mathbf{\Theta}$ should be found. Assume that $N_b=R_cSLJ$ information bits are transmitted, then $J$ coded bits are transmitted by each of the $SL$ parallel subchannels. Hence, $PJ$ coded bits are transmitted by a precoded subchannel set. Note that $N_b$ information bits can provide $2^{N_b}$ different bit codewords. Hence, if $PJ$ is smaller than $N_b$, there always exists at least a pair of bit codewords whose $PJ$ coded bits transmitted by a precoded subchannel set are the same. The reason is that the total possible number of bit sequences for a precoded subchannel set, which is $2^{PJ}$, is smaller than the total possible bit codewords $2^{N_b}$. As a result, the precoded subchannel set is non-effective. Therefore, $PJ$ cannot be smaller than $N_b$, which implies that $P \geq R_cSL$. Since $P$ is an integer, the minimum effective dimension of $\mathbf{\Theta}$ is $P=\lceil R_cSL \rceil$. Note that $P=\lceil R_cSL \rceil$ is only proved in this subsection to be a necessary condition because the requirement, i.e., at least an errored bit of each error event is transmitted through each precoded subchannel set, is also related to the convolutional code and the bit interleaver.  

\subsection{Minimum Effective Number of Precoding Subchannel Sets} \label{subsec:pre_num}
Assume that at least an errored bit of each error event is transmitted through each precoded subchannel by a properly designed combination of the convolutional code and the bit interleaver. Then, every precoded subchannel set is effective. However, it still does not guarantee full diversity. Note that the full diversity condition derived in Section \ref{sec:diversity} requires that $\rho_{l,1} \neq 0, \, \forall l$ of all error events. It is also illustrated in Section \ref{subsec:pre_dim} that the non-precoded subchannels result in the same weights for both precoded and non-precoded BICMB-OFDM-SG. As a result, if a first subchannel of a subcarrier is not precoded, there always exists at least one error path with no errored bits transmitted through that subchannel when $R_cSL > 1$, as proved in \cite{Li_DA_BICMB_OFDM}. In that case, full diversity cannot be achieved even if all precoded subchannel sets are effective. Therefore, the first subchannels of all subcarriers should be precoded. Since there are $L$ subcarriers which offer $L$ first subchannels in total, and each $\mathbf{\Theta}$ can precode $P \geq \lceil R_cSL \rceil$ subchannels, then the minimum effective number of precoding subchannel sets is $N_p=\lceil L/P \rceil$.   

The aforementioned full diversity requirement is that the first subchannels of all subcarriers should be precoded. However, if $\lceil L/P \rceil P > SL$, the full diversity requirement cannot be satisfied because not all first subchannels of all subcarriers can be precoded by $\mathbf{\Theta}$ with effective dimension $P \geq \lceil R_cSL \rceil$. In fact, the case of $\lceil L/P \rceil P > SL$ can only happen when $S=1$. In other words, when $S \geq 2$, $\lceil L/P \rceil P \leq SL$ is always valid, which is proved in the following.
\begin{IEEEproof}
Note that $\left\lceil L/P \right\rceil \geq 1$. If $\lceil L/P \rceil = 1$, $\lceil L/P \rceil P \leq SL$ is always valid because $P \leq SL$. On the other hand, if $\lceil L/P \rceil \geq 2$, then $P<L$. Because $S \geq 2$, then
\begin{align}
\lceil L/P \rceil P < 2L \leq SL.
\label{eq:pre_dim_proof}
\end{align}
This concludes the proof.
\end{IEEEproof}

As a result, when $S \geq 2$, the minimum effective number of $\mathbf{\Theta}$ is $N_p = \lceil L/P \rceil$ with $P = \lceil R_cSL \rceil$. Similar selection of $N_p$ and $P$ can be applied for $S=1$ if $N_pP \leq L$. Otherwise, if $N_pP > L$ for $S=1$, the dimension $P$ of $\mathbf{\Theta}$ needs to be increased so that the product of $P$ and its corresponding $N_p$ satisfies $N_pP \leq L$.
\\\textbf{Example:}
Consider the parameters $N_t=N_r=2$, $L=4$, $S=1$, and $R_c=2/3$, then the minimum effective dimension of $\mathbf{\Theta}$ is $P = \lceil R_cSL \rceil = 3$. Hence, the minimum effective number of precoding subchannel sets is $N_p = \lceil L/P \rceil = 2$. As a result, $N_pP=6$ subchannels are required to be precoded by two sets. However, there are only four subchannels. Therefore, full diversity cannot be achieved with the selection of $P=3$ and $N_p=2$. Hence, $P=4$ should be applied instead. In that case, $N_p=1$, and $N_pP=4\leq L$. Therefore, $P=4$ and $N_p=1$ is the minimum effective selection. 

Note that similar to the discussion on $P$ in Section \ref{subsec:pre_dim}, the minimum effective selection of $P$ and its corresponding $N_p$ is only a necessary condition, because the convolutional code and the interleaver also need to be considered to satisfy the requirement, i.e., at least one errored bit of each error event is transmitted through each precoded subchannel set.  

\subsection{Selection of Precoded Subchannels} \label{subsec:pre_select}
According to (\ref{eq:pep_diversity}), (\ref{eq:diversity}), (\ref{eq:weight_lambda_beta}), and (\ref{eq:beta}), the diversity of BICMB-OFDM-SG with precoding also depends on the $\alpha$-spectra of BICMB-OFDM-SG without precoding. In fact, the $\alpha$-spectra are related with the bit interleaver and the trellis structure of the convolutional code, and are independent of the precoding matrix.  Note that the $\alpha$-spectra can be derived by a similar approach to BICMB in the case of flat fading MIMO channels presented in \cite{Park_DA_BICMB}, or by computer search. 
Based on the $\alpha$-spectra for a certain combination of the convolutional code and the bit interleaver, the selection of precoded subchannels should be properly designed in order to satisfy the condition of $\rho_{l,1} \neq 0, \, \forall l$ for all error events.
\\\textbf{Example:}
Consider the $4$-state $R_c=1/2$ convolutional code with generator polynomials $(5, 7)$ in octal representation, in a subcarrier group of BICMB-OFDM-SG without precoding with parameters $N_t=N_r=S=L=2$ and $M=64$. Two types of spatial interleavers are considered to demonstrate the way to select precoded subchannels for each set. The $1$st, $2$nd, $3$rd, and $4$th subchannels are symbolically represented as $a$, $b$, and $c$ and $d$, respectively. The spatial interleaver used in $\mathcal{T}_1$ is a simple bit-by-bit rotating switch on four subchannels. For $\mathcal{T}_2$, the spatial interleaver is simply rotated $6$-bits-by-$6$-bits on four subchannels. In the following transfer functions, each term represents an $\alpha$-spectrum, and the exponents of $a$, $b$, $c$, and $d$ of a term indicate its corresponding values of the $\alpha$-spectrum.
\ifCLASSOPTIONonecolumn
\begin{align}
\mathbf{T}_1 &= Z^5 (a^2 b^2 d + b c^2 d^2) + \nonumber\\
&\quad\,\, Z^6 (a^2 b^2 d^2 + 2 a^2 b c^2 d + b^2 c^2 d^2) + \nonumber\\
&\quad\,\, Z^7 (a^2 b^3 c^2 + 2 a^2 b^2 c^2 d + a^2 b^2 d^3 + 2 a^2 b c^2 d^2 + a^2 c^2 d^3 + b^3 c^2 d^2) + \nonumber\\
&\quad\,\, Z^8 (a^4 b^2 c^2 + 4 a^2 b^3 c^2 d + 4 a^2 b^2 c^2 d^2 + a^2 b^2 d^4 + 4 a^2 b c^2 d^3 + a^2 c^4 d^2 + b^4 c^2 d^2) + \cdots . \\
\mathbf{T}_2 &= Z^5 (a^5 + a^3b^2 + a^3d^2 + a^2b^3 + a^2d^3 + b^5 + b^3c^2 + b^2c^3 + c^5 + c^3d^2 + c^2d^3 + d^5) +  \nonumber\\
&\quad\,\, Z^6 (a^4b^2 + a^4d^2 + 3a^3b^3 + 3a^3d^3 + a^2b^4 + a^2b^2c^2 + a^2b^2d^2 + a^2c^2d^2 + a^2d^4 + b^4c^2 + 3b^3c^3 + \nonumber \\ 
&\qquad b^2c^4 + b^2c^2d^2 + c^4d^2 + 3c^3d^3 + c^2d^4) + \nonumber\\
&\quad\,\, Z^7 (2a^4b^3 + 2a^4d^3 + 2a^3b^4 + 3a^3b^2c^2 + 2a^3b^2d^2 + a^3bc^3 + a^3c^3d + 2a^3c^2d^2 + 2a^3d^4 + 2a^2b^3c^2 + \nonumber \\
&\qquad 2a^2b^3d^2 + 2a^2b^2c^3 + 3a^2b^2d^3 + 3a^2c^3d^2 + 2a^2c^2d^3 + ab^3d^3 + 2b^4c^3 + 2b^3c^4 + 3b^3c^2d^2 + \nonumber \\ &\qquad b^3cd^3 + 2b^2c^3d^2 + 2b^2c^2d^3 + 2c^4d^3 + 2c^3d^4) + \cdots .
\label{eq:transfer_example_diffdemux}
\end{align}
\else
\begin{align}
\mathbf{T}_1 &= Z^5 (a^2 b^2 d + b c^2 d^2) + \nonumber\\
&\quad\,\, Z^6 (a^2 b^2 d^2 + 2 a^2 b c^2 d + b^2 c^2 d^2) + \nonumber\\
&\quad\,\, Z^7 (a^2 b^3 c^2 + 2 a^2 b^2 c^2 d + a^2 b^2 d^3 + 2 a^2 b c^2 d^2 + \nonumber \\ 
&\qquad a^2 c^2 d^3 + b^3 c^2 d^2) + \nonumber\\
&\quad\,\, Z^8 (a^4 b^2 c^2 + 4 a^2 b^3 c^2 d + 4 a^2 b^2 c^2 d^2 + a^2 b^2 d^4 + \nonumber \\
&\qquad 4 a^2 b c^2 d^3 + a^2 c^4 d^2 + b^4 c^2 d^2 ) + \cdots . \\
\mathbf{T}_2 &= Z^5 (a^5 + a^3b^2 + a^3d^2 + a^2b^3 + a^2d^3 + b^5 + b^3c^2 + \nonumber \\
&\qquad b^2c^3 + c^5 + c^3d^2 + c^2d^3 + d^5) +  \nonumber\\
&\quad\,\, Z^6 (a^4b^2 + a^4d^2 + 3a^3b^3 + 3a^3d^3 + a^2b^4 + a^2b^2c^2 + \nonumber \\
&\qquad a^2b^2d^2 + a^2c^2d^2 + a^2d^4 + b^4c^2 + 3b^3c^3 + \nonumber \\ 
&\qquad b^2c^4 + b^2c^2d^2 + c^4d^2 + 3c^3d^3 + c^2d^4) + \nonumber\\
&\quad\,\, Z^7 (2a^4b^3 + 2a^4d^3 + 2a^3b^4 + 3a^3b^2c^2 + 2a^3b^2d^2 + \nonumber \\ 
&\qquad a^3bc^3 + a^3c^3d + 2a^3c^2d^2 + 2a^3d^4 + 2a^2b^3c^2 + \nonumber \\
&\qquad 2a^2b^3d^2 + 2a^2b^2c^3 + 3a^2b^2d^3 + 3a^2c^3d^2 + \nonumber \\
&\qquad 2a^2c^2d^3 + ab^3d^3 + 2b^4c^3 + 2b^3c^4 + 3b^3c^2d^2 + \nonumber \\ 
&\qquad b^3cd^3 + 2b^2c^3d^2 + 2b^2c^2d^3 + 2c^4d^3 + 2c^3d^4) + \cdots .
\label{eq:transfer_example_diffdemux}
\end{align}
\fi
Without precoding, to satisfy the full diversity condition $\alpha_{l,1} \neq 0, \, \forall l$ of all error events, each term in the transfer function should include both $a$ and $c$. For $\mathbf{T}_1$, the $\alpha$-spectra $\mathbf{A}=[0 \, 1; 2 \, 2]$ and $\mathbf{A}=[2 \, 2; 0 \, 2]$ without full diversity dominate the performance. Since each term of the transfer function includes at least $a$ or $c$ for each term, a two-dimensional $\mathbf{\Theta}$ precoding $a$ and $c$ can satisfy the full diversity condition $\rho_{l,1} \neq 0, \, \forall l$ of all error events, for which the selection of $P=\lceil R_cSL \rceil=2$ and $N_p=\lceil L/P \rceil=1$ is the minimum effective choice derived in Section \ref{subsec:pre_dim} and Section \ref{subsec:pre_num}. On the other hand, for $\mathbf{T}_2$, the $\alpha$-spectra $\mathbf{A}=[0 \, 5; 0 \, 0]$ and $\mathbf{A}=[0 \, 0; 0 \, 5]$ without full diversity dominate the performance. Since the transfer function also includes $\alpha$-spectra $\mathbf{A}=[5 \, 0; 0 \, 0]$ and $\mathbf{A}=[0 \, 0; 5 \, 0]$, a four-dimensional $\mathbf{\Theta}$ precoding all subchannels is required to provide full diversity, for which the selection of $P=\lceil R_cSL \rceil=4$ and $N_p=\lceil L/P \rceil=1$ is not the minimum effective choice derived in Section \ref{subsec:pre_dim} and Section \ref{subsec:pre_num}.

The aforementioned example shows that the minimum effective selection of $P$ and $N_p$ may not be effective when the bit interleaver is not properly designed. As a result, the precoded subchannels and the bit interleaver should be jointly designed to provide full diversity. In Section \ref{subsec:pre_dim} and Section \ref{subsec:pre_num}, the minimum effective selection of $P$ and $N_p$ is provided as a necessary full diversity condition. In the following, the minimum effective selection of $P$ and $N_p$ is proved to be sufficient to provide full diversity with the joint design of precoded subchannels and the bit interleaver. 

\begin{IEEEproof}
Consider the rate of the convolutional code $R_c = k_c/n_c$ where $k_c$ and $n_c$ are positive integers with $k_c<n_c$, which implies that each $k_c$ branches in the trellis of convolutional coded generates $n_c$ coded bits. If the spatial de-multiplexer is not a random switch for the whole packet, the period of the spatial de-multiplexer is an integer multiple of the Least Common Multiple (LCM) of $n_c$ and $SL$. Note that a period of the interleaver is restricted to an integer multiple of trellis branches. Define $Q$ = $LCM(n_c, SL)$ as the number of coded bits for a minimum period, which is considered below. Since each subchannel needs to be evenly employed for a period, $Q/(SL)$ coded bits are assigned on each subchannel. Therefore, $QP/(SL)$ coded bits are transmitted through one precoded subchannel set, which offer the same effect on the diversity. Note that the trellis structure of convolutional code can be designed such that the coded bits generated from the first branch splitting from the zero state are all errored bits of an error event. Consequently, to guarantee $\rho_{l,1} \neq 0, \, \forall l$ of all error events, it is sufficient to consider only the first branches that split from the zero state in one period because of the repetition property of convolutional code. In other words, if at least one coded bit for each precoded subchannel set is assigned for each branch, full diversity is achieved. Note that there are $QR_c$ branches in a minimum period. Because $P\geq R_cSL$, then $QP/(SL)\geq QR_c$. As a result, all branches in a minimum period can be assigned at least one coded bit transmitted through each precoded subchannel set, which guarantees full diversity. 

This concludes the proof.
\end{IEEEproof}

\subsection{Complexity} \label{subsec:pre_complexity}
With precoding, BICMB-OFDM-SG without the full diversity restriction of $R_cSL \leq 1$ can achieve full diversity with the trade-off of an increased decoding complexity. Assume that square QAM with constellation size $N_m$ is employed. Specifically, the complexity of ML metric calculation for (\ref{eq:ml_bit_metrics}) depends on only one of the real and imaginary parts corresponding to the coded bit \cite{Akay_BICM_LCD, Akay_LCD_BICM}. If quantization is applied, the complexity is proportional to $1$, denoted by $\mathcal{O}(1)$. With precoding matrices introduced in \cite{Park_CPB}, the worst-case complexity of ML metric calculation for the precoded bits in (\ref{eq:ml_bit_metrics_decomposed}) is $\mathcal{O}(N_m^{P-1})$ by using a real-valued Sphere Decoding (SD) based on the real lattice representation in \cite{Azzam_SD_NLR, Azzam_SD_RLR}, plus quantization of the last two layers. Since the complexity for ML metric calculation of the precoded part dominates the overall complexity, the ML decoding complexity of precoded BICMB-OFDM-SG is considered as $\mathcal{O}(N_m^{P-1})$ for the worst case.

In \cite{Li_GCMB, Li_BICMB_PC, Li_MB_PC}, PSTBCs, which have the properties of full rate, full diversity, uniform average transmitted energy per antenna, good shaping of the constellation, and nonvanishing constant minimum determinant for increasing spectral efficiency which offers high coding gain, have been considered as an alternative scheme to replace the constellation precoding technique for both uncoded and coded SVD beamforming with constellation precoding in the case of flat fading MIMO channels. By doing so, the decoding complexity in dimensions $2$ and $4$ can be reduced while the performance is almost the same. The reason of the complexity reduction is that, due to the special property of the generation matrices in dimensions $2$ and $4$, the real and imaginary parts of the received signal can be separated, and only the part corresponding to the coded bit is required to calculate one bit metric for Viterbi decoder. For BICMB-OFDM-SG, PSTBCs can also be applied. As a result, the worst-case decoding complexity of ML metric calculation for the precoded bits in (\ref{eq:ml_bit_metrics_decomposed}) is $\mathcal{O}(N_m^{0.5})$ for $P=2$ and $\mathcal{O}(N_m^{1.5})$ for $P=4$. Therefore, the ML decoding complexity of precoded BICMB-OFDM-SG is considered as $\mathcal{O}(N_m^{0.5})$ for $P=2$ and $\mathcal{O}(N_m^{1.5})$ for $P=4$ in the worst case.

Table \ref{table:complexity} summarizes the worst-case ML decoding complexity of BICMB-OFDM-SG with different dimensions of precoding matrices when square $N_m$-QAM is employed. 
\begin{table}[!t]
\renewcommand{\arraystretch}{1.3}
\caption{Worst-case ML decoding complexity of precoded BICMB-OFDM-SG.}
\centering
\begin{tabular}{|c|c|c|}
\hline
& \cite{Park_CPB} & \cite{Li_GCMB, Li_BICMB_PC, Li_MB_PC} \\
\hline
$P=2$ & $\mathcal{O}(N_m)$ & $\mathcal{O}(N_m^{0.5})$ \\
\hline
$P=3$ & $\mathcal{O}(N_m^{2})$ & N/A \\
\hline
$P=4$ & $\mathcal{O}(N_m^{3})$ & $\mathcal{O}(N_m^{1.5})$ \\
\hline
$P\geq 5$ & $\mathcal{O}(N_m^{P-1})$ & N/A \\
\hline
\end{tabular}
\label{table:complexity}
\end{table}
Note that PSTBCs are only available in dimensions $2$, $3$, $4$, and $6$, and dimensions $3$ and $6$ have no complexity advantage and do not employ QAM. Table \ref{table:complexity} shows that the complexity of $P=4$ can be lower than $P=3$ by employing PSTBC. As a result, if the minimum effective dimension of the precoding matrices is $P=3$, then $P=4$ and its corresponding $N_p$ should be applied if they are a valid selection. 

In the sequel, PSTBCs are incorporated into a design algorithm, and in Section \ref{sec:results}, simulation results are provided with systems employing PSTBCs.

\subsection{Full-Diversity Precoding Design Summary} \label{subsec:pre_design}
Based on the discussion of the previous subsections, a sufficient method of the full-diversity precoding design for BICMB-OFDM-SG with $R_cSL>1$ is summarized as the following steps.
\begin{enumerate}
\item Calculate $P=\lceil R_cSL \rceil$. Set $\mathrm{flag}=0$. 
\item If $P=3$ and $\mathrm{flag}=0$, set $P=4$. If $P=3$ and $\mathrm{flag}=1$, set $P=5$. Otherwise, go to 3).
\item Calculate $N_p=\lceil L/P \rceil$.
\item Calculate $N_pP$. If $N_pP>SL$ and $P=3$, go to 2). If $N_pP>SL$, and $P=4$, set $P=3$ and $\mathrm{flag}=1$, then go to 3). If $N_pP>SL$ and $P \neq 3$ and $P \neq 4$, set $P=P+1$ and go to 2). Otherwise, go to 5).
\item If $P=2$ or $P=4$, PSTBCs are applied as in \cite{Li_GCMB, Li_BICMB_PC, Li_MB_PC}. Otherwise, constellation precoding is applied with precoding matrices introduced in \cite{Park_CPB}.
\item Select $N_pP$ precoded subchannels which include all the $L$ first subchannels of all subcarriers.
\item Design a bit interleaver pattern of $Q$ = $LCM(n_c, SL)$ coded bits for a period by assigning one precoded subchannel from each set to each branch.
\end{enumerate}

\subsection{Discussion}
In this paper, the dimension $P$ of $\mathbf{\Theta}$ for each precoded subchannel set is assumed to be the same, which is actually not necessary. Applying precoded subchannel sets with different number of subchannels may achieve lower decoding complexity because the complexity increases exponentially as the dimension increases, as shown in Table \ref{table:complexity}.
\\\textbf{Example:}
Consider the parameters $N_t=N_r=2$, $L=7$, $S=1$, and $R_c=1/3$, then $P=7$ and $N_p=1$ is the minimum effective selection if different dimensions of the precoded subchannel sets are not considered. On the other hand, if different dimensions of the precoded subchannel sets are considered, two sets with $P_1=3$ and $P_2=4$ can also provide full diversity, which can achieve lower decoding complexity.

In this paper, the number of employed subchannels by SVD for each subcarrier is assumed to be the same, which is $S$. However, they could be different in practice. In that case, the full diversity condition for one subcarrier group of BICMB-OFDM-SG without precoding is $R_c\sum_{l=1}^{L}S_l \leq 1$ where $S_l$ denotes the number of employed subchannels by SVD for the $l$th subcarrier of the group
. With precoding, the minimum effective selection of $P\geq\lceil R_c\sum_{l=1}^{L}S_l \rceil$ and $N_p=\lceil L/P \rceil$ can be derived applying the same method as summarized in Section \ref{subsec:pre_dim} and Section \ref{subsec:pre_num} respectively for BICMB-OFDM-SG with $R_c\sum_{l=1}^{L}S_l > 1$. In fact, if $\sum_{l=1}^{L}S_l<N_pP\leq \min\{N_t,N_r\}L$, instead of retrying different selections of $P$ and $N_p$ with higher decoding complexity, the current selection can become valid by increasing the number of employed subchannels at each subcarrier so that $\sum_{l=1}^{L}S_l=N_pP$. In such a way, the minimum decoding complexity can be achieved with increased number of employed parallel subchannels.

As shown in Table \ref{table:complexity}, the worst-case ML decoding complexity of precoded BICMB-OFDM-SG is $\mathcal{O}(N_m^{0.5})$, $\mathcal{O}(N_m^{1.5})$, and $\mathcal{O}(N_m^{2})$ for dimensions $2$, $4$, and $3$, respectively. Note that the worst-case complexity of $\mathcal{O}(N_m^{4})$ for $P=5$ has a significant increase. As a result, if $P\geq5$, instead of applying the precoding directly, employing a convolutional code with smaller rate $R_c$ so that the minimum effective dimension $P\leq4$ may be a more reasonable option. 
 
In \cite{Park_BICMB_CP, Park_MB_CP, Park_CPMB}, constellation precoding is applied to BICMB without the full diversity restriction $R_cS\leq1$ for flat fading MIMO channels. It was presented that partial precoding could achieve both full diversity and full multiplexing with the properly designed combination of the convolutional code, the bit interleaver, and the constellation precoder. However, the general full-diversity precoding design was not provided. Since BICMB of flat fading MIMO channels can be considered as a subcarrier of BICMB-OFDM-SG in the frequency domain with $L=1$, the full-diversity precoding design proposed in this paper can be applied to BICMB with $R_cS>1$ for flat fading MIMO channels.
\section{Simulation Results} \label{sec:results}

To verify the diversity analysis and the full-diversity precoding design, $2 \times 2$ $M=64$ BICMB-OFDM-SG with $L=2$ and $L=4$ using $4$-QAM, as well as $4 \times 4$ $M=64$ BICMB-OFDM-SG with $L=2$ using $4$-QAM, are considered for simulations. The number of employed subchannels for each subcarrier and the dimension of precoding matrix for each precoded subchannel set are assumed to be the same, respectively. The generator polynomials in octal for the convolutional codes with $R_c=1/4$ and $R_c= 1/2$ are $(5,7,7,7)$ and $(5,7)$ respectively, and the codes with $R_c = 2/3$ and $R_c = 4/5$ are punctured from the $R_c=1/2$ code \cite{Haccoun_PCC}. The length of CP is $L_{cp}=16$. Each OFDM symbol has $4\mu \mathrm{s}$ duration, of which $0.8\mu \mathrm{s}$ is CP. Equal power channel taps are considered. The bit interleaver employs simple rotation for BICMB-OFDM-SG with non-effective precoding selections and without precoding. For BICMB-OFDM-SG with effective precoding selections, the proposed full-diversity precoding design in this paper is employed. In the figures, NP indicates non-precoded. Note that unequal power channel taps are not considered because they do not affect the maximum achievable diversity as discussed in \cite{Li_DA_BICMB_OFDM}. Also note that simulations of $2 \times 2$ BICMB-OFDM with $L=2$ and $L=4$ as well as $4 \times 4$ BICMB-OFDM with $L=2$ are shown in this section because the diversity values could be investigated explicitly through figures. In our simulations, we will show that the maximum diversity of $N_rN_tL$ \cite{Jafarkhani_STC} is achieved.

\subsection{Non-Precoded BICMB-OFDM-SG} \label{subsec:results_np}
\ifCLASSOPTIONonecolumn
\begin{figure}[!t]
\centering \includegraphics[width = 1.0\linewidth]{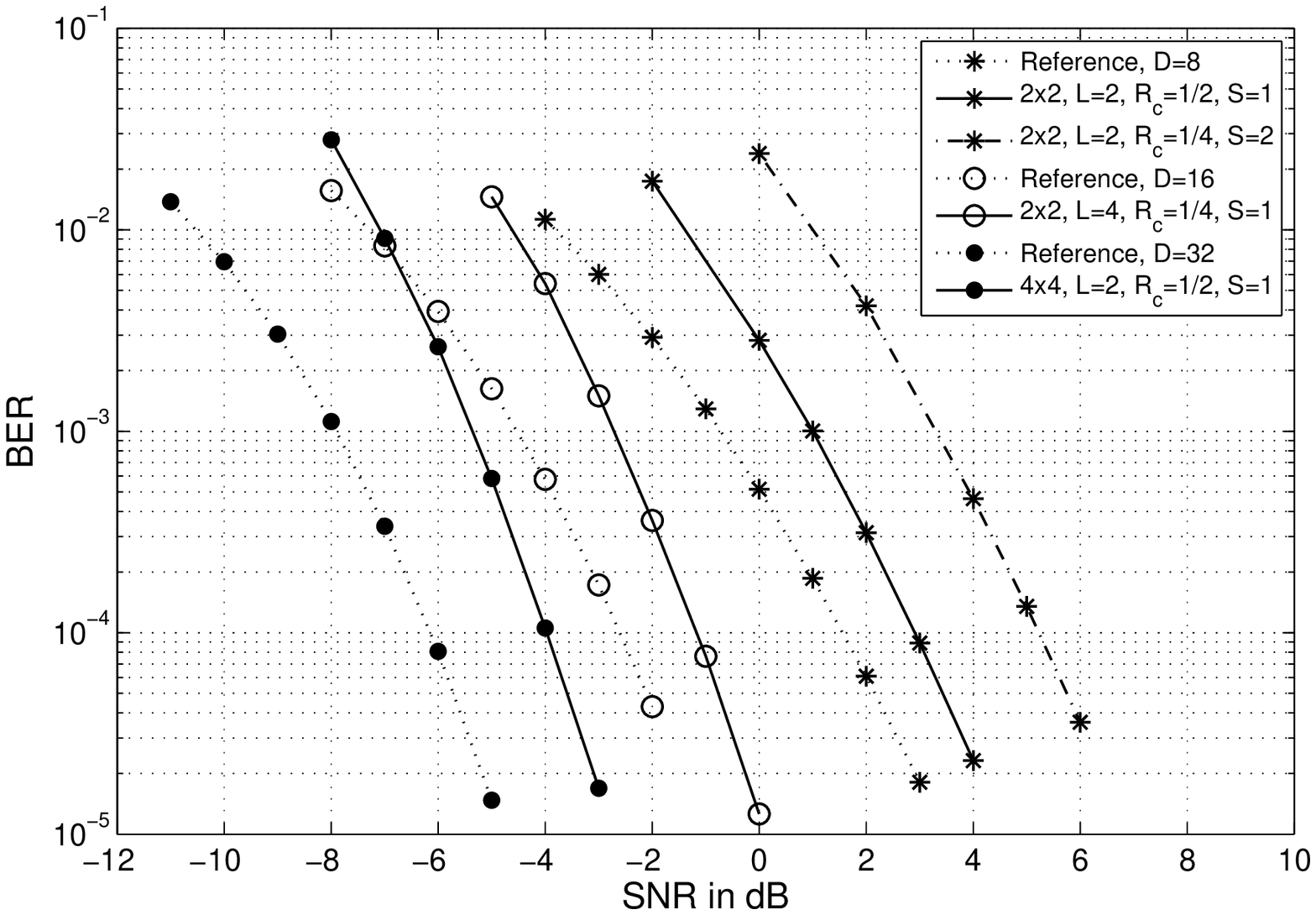}
\caption{BER vs. SNR for non-precoded BICMB-OFDM-SG achieving full diversity.}
\label{fig:np_div}
\end{figure}
\else
\begin{figure}[!t]
\centering \includegraphics[width = 1.0\linewidth]{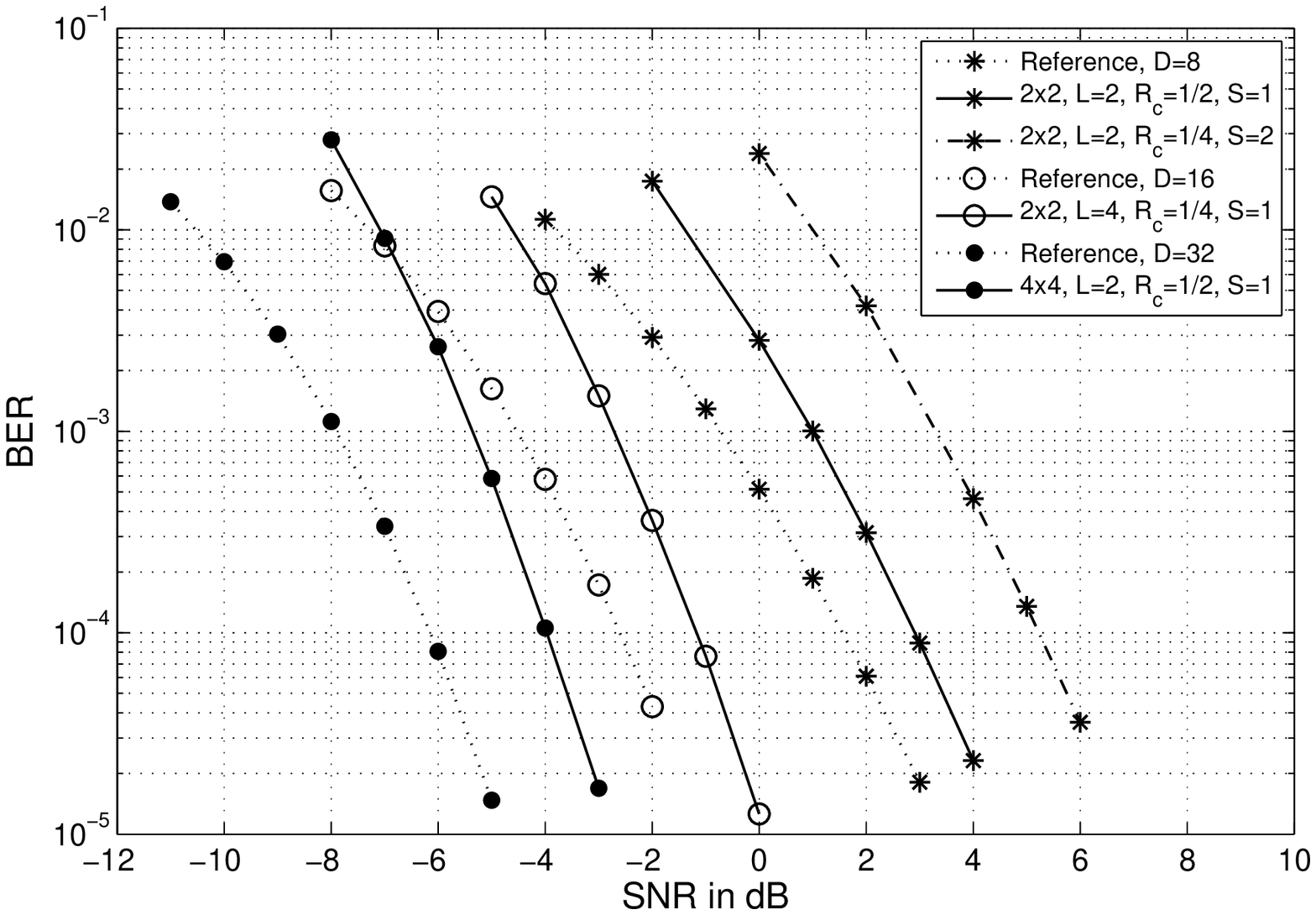}
\caption{BER vs. SNR for non-precoded BICMB-OFDM-SG achieving full diversity.}
\label{fig:np_div}
\end{figure}
\fi

Fig. \ref{fig:np_div} shows the Bit Error Rate (BER) performance of non-precoded BICMB-OFDM-SG achieving full diversity for different system parameters. System configurations of $2 \times 2$ $L=2$ $R_c=1/2$ $S=1$, $2 \times 2$ $L=2$ $R_c=1/4$ $S=2$, $2 \times 2$ $L=4$ $R_c=1/4$ $S=1$, and $4 \times 4$ $L=2$ $R_c=1/2$ $S=1$ are considered. According to \cite{Li_DA_BICMB_OFDM}, they all achieve their corresponding full diversity orders because they all satisfy the full diversity condition of $R_cSL\leq1$ for non-precoded BICMB-OFDM-SG. The full diversity orders of $2 \times 2$ $L=2$, $2 \times 2$ $L=4$, and $4 \times 4$ $L=2$ systems are $8$, $16$, and $32$ respectively. In the figure, the theoretical probability of error for Maximum Ratio Combining (MRC) diversity systems with $N_r=D\in\{8,16,32\}$ receive antennas using Binary Phase-Shift Keying (BPSK) over Rayleigh flat fading channels are drawn as references to the cases of diversity orders $D$ \cite{Barry_DC}. Note that Fig. \ref{fig:np_div} provides full diversity references for this section. Since this paper focuses on full diversity, references for the non-full diversity orders are not offered in figures. Note that the non-full diversity orders in this section are derived by the results of (\ref{eq:pep_diversity}) and (\ref{eq:diversity}) as discussed in Section \ref{sec:diversity}.        

\subsection{$2 \times 2$ $L=2$ Precoded BICMB-OFDM-SG} \label{subsec:results_2taps_2x2}
\ifCLASSOPTIONonecolumn
\begin{figure}[!t]
\centering \includegraphics[width = 1.0\linewidth]{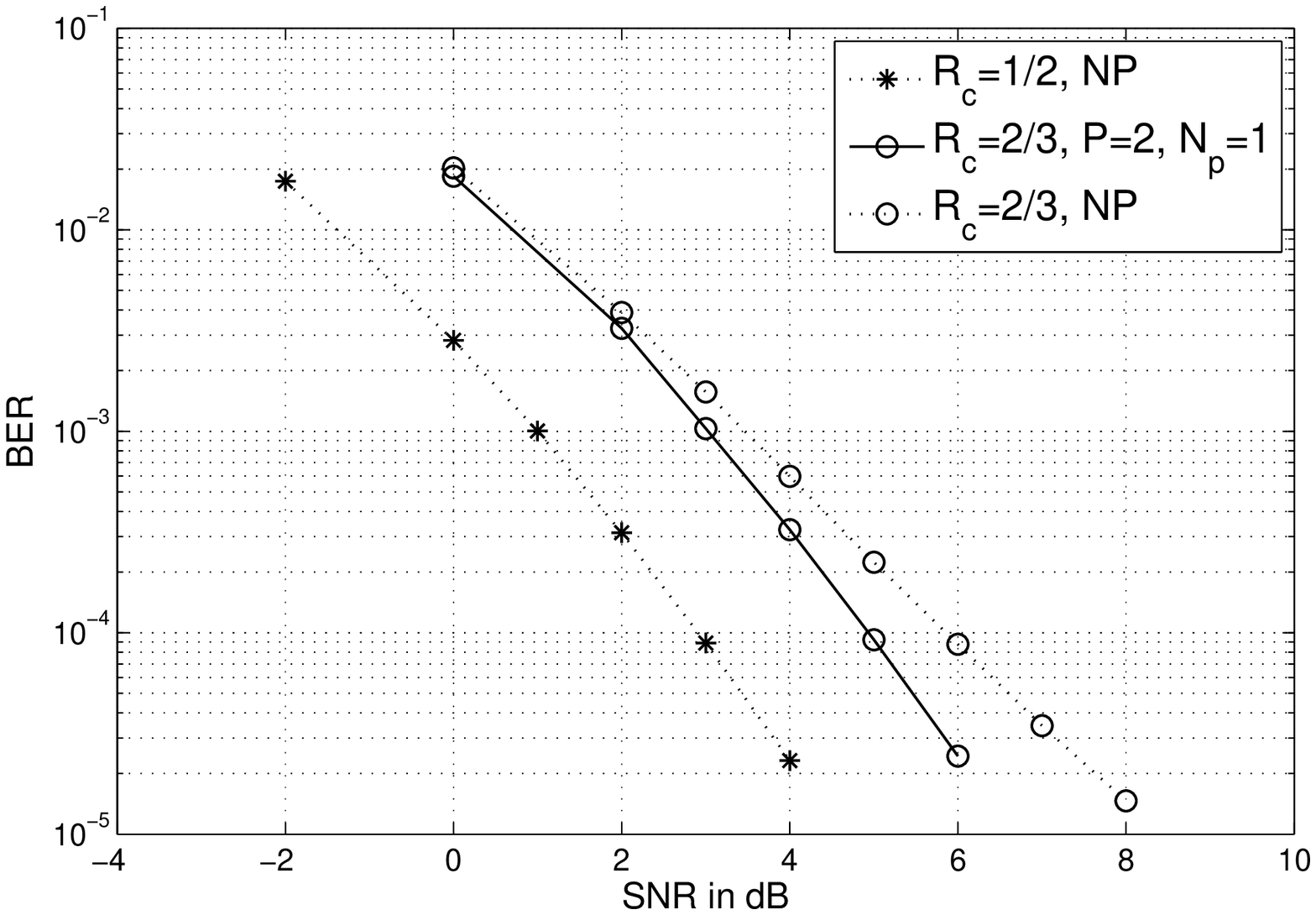}
\caption{BER vs. SNR for $2 \times 2$ $L=2$ $M=64$ $S=1$ BICMB-OFDM-SG with and without precoding.}
\label{fig:pre_2L1S2x2}
\end{figure}
\else
\begin{figure}[!t]
\centering \includegraphics[width = 1.0\linewidth]{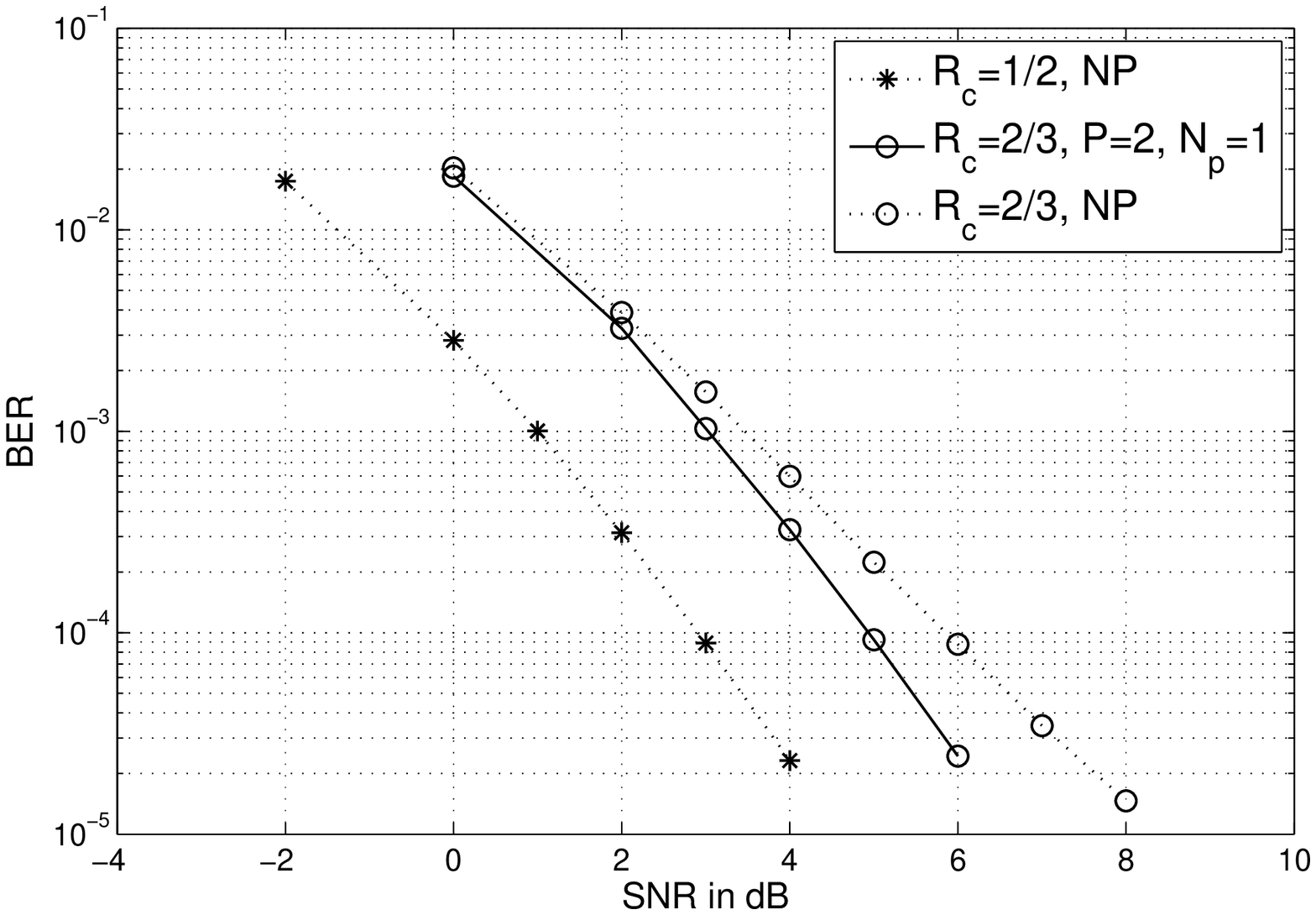}
\caption{BER vs. SNR for $2 \times 2$ $L=2$ $M=64$ $S=1$ BICMB-OFDM-SG with and without precoding.}
\label{fig:pre_2L1S2x2}
\end{figure}
\fi

Fig. \ref{fig:pre_2L1S2x2} shows the BER performance of $2 \times 2$ $L=2$ $M=64$ $S=1$ BICMB-OFDM-SG with and without precoding for different $R_c$. For $R_c=1/2$, full diversity of $8$ can be achieved even if precoding is not applied as shown in Fig \ref{fig:np_div}, because $R_cSL \leq 1$ \cite{Li_DA_BICMB_OFDM}. On the other hand, in the case of $R_c=2/3$, the diversity order is $4$ instead of full diversity since $R_cSL > 1$. However, with the full-diversity precoding design proposed in this paper of $P=2$ and $N_p=1$, full diversity of $8$ is successfully recovered.

\ifCLASSOPTIONonecolumn
\begin{figure}[!t]
\centering \includegraphics[width = 1.0\linewidth]{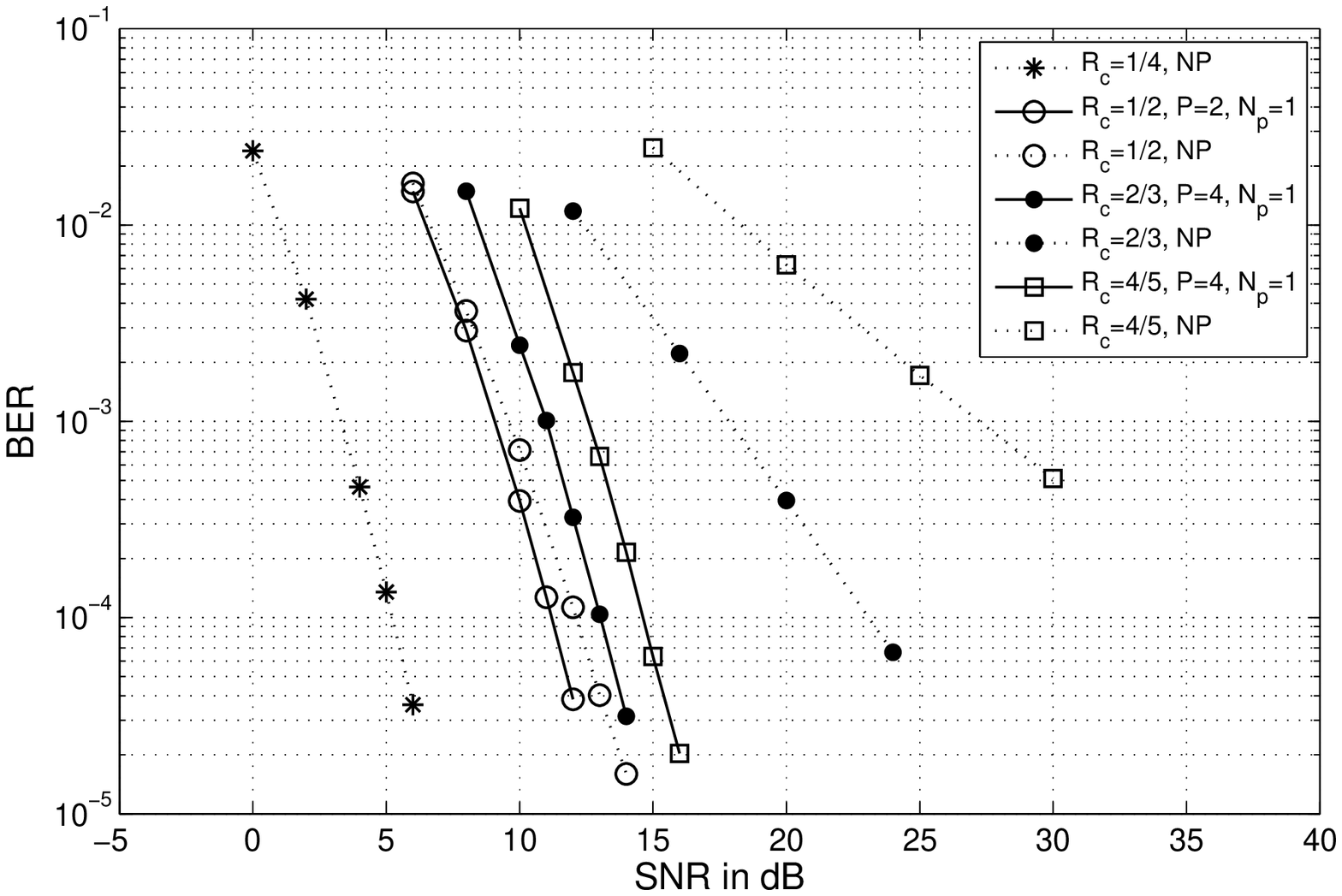}
\caption{BER vs. SNR for $2 \times 2$ $L=2$ $M=64$ $S=2$ BICMB-OFDM-SG with and without precoding.}
\label{fig:pre_2L2S2x2}
\end{figure}
\else
\begin{figure}[!t]
\centering \includegraphics[width = 1.0\linewidth]{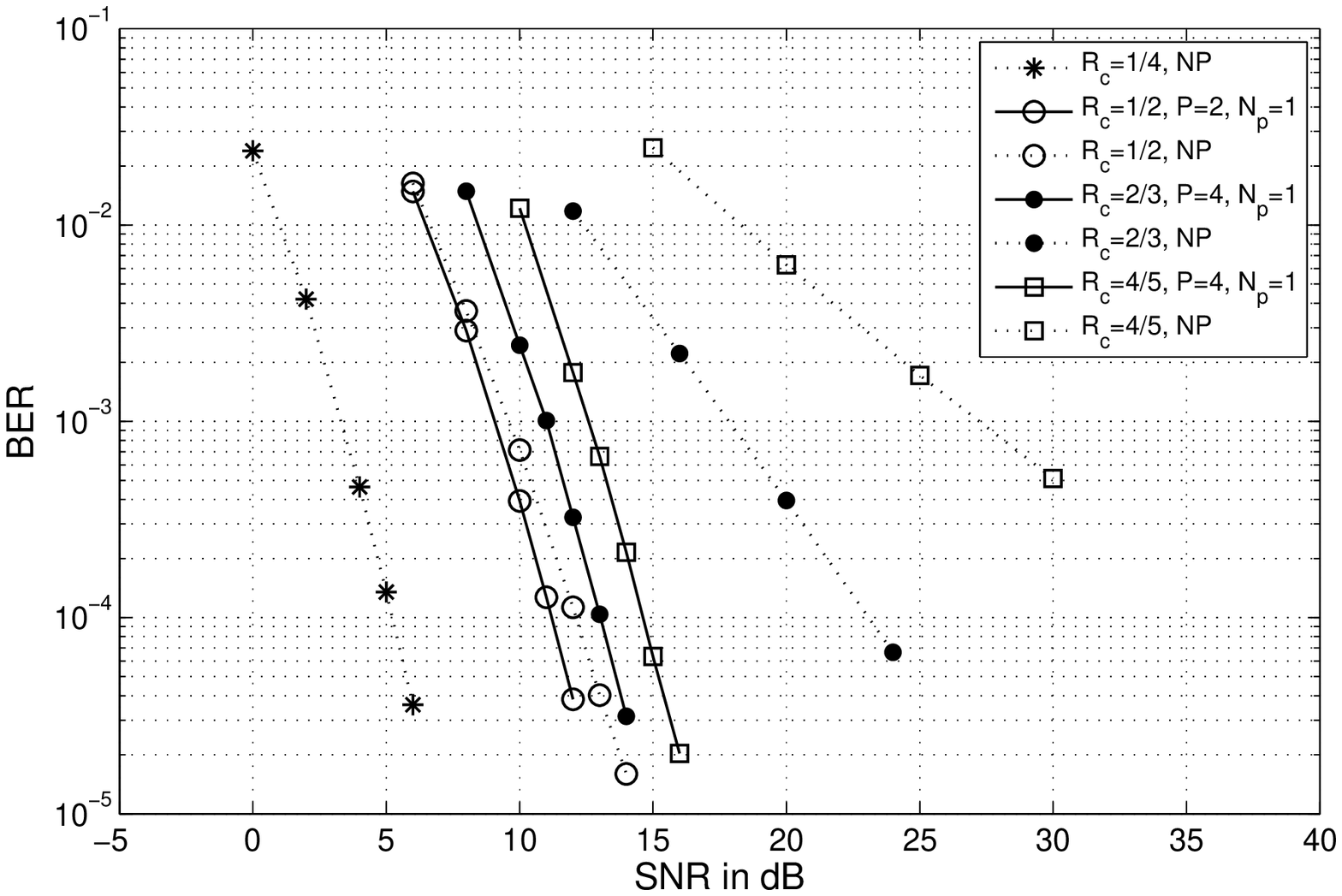}
\caption{BER vs. SNR for $2 \times 2$ $L=2$ $M=64$ $S=2$ BICMB-OFDM-SG with and without precoding.}
\label{fig:pre_2L2S2x2}
\end{figure}
\fi

Similarly, Fig. \ref{fig:pre_2L2S2x2} shows the BER performance of $2 \times 2$ $L=2$ $M=64$ $S=2$ BICMB-OFDM-SG with and without precoding for different $R_c$. For $R_c=1/4$, full diversity of $8$ can be achieved even without precoding as shown in Fig \ref{fig:np_div}, since $R_cSL \leq 1$ \cite{Li_DA_BICMB_OFDM}. On the other hand, in the cases of $R_c=1/2$, $R_c=2/3$, and $R_c=4/5$, the diversity orders are $5$, $2$, and $1$ respectively, and the full diversity degradations result from $R_cSL > 1$. Nevertheless, full diversity of $8$ can be restored by employing the full-diversity precoding design proposed in this paper. The corresponding selections are $P=2$ $N_p=1$, $P=4$ $N_p=1$, and $P=4$ $N_p=1$ for $R_c=1/2$, $R_c=2/3$, and $R_c=4/5$, respectively.

\ifCLASSOPTIONonecolumn
\begin{figure}[!t]
\centering \includegraphics[width = 1.0\linewidth]{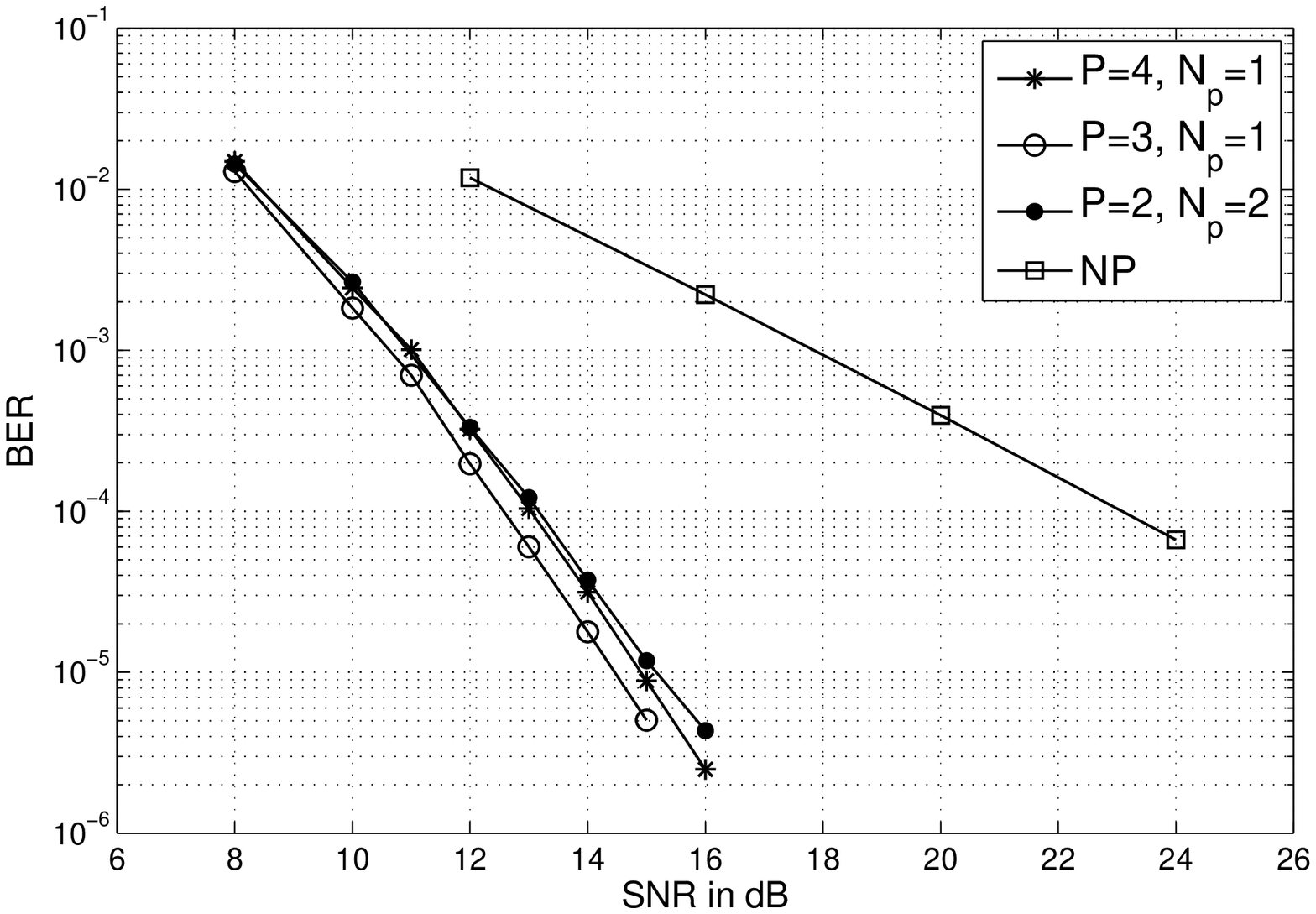}
\caption{BER vs. SNR for $2 \times 2$ $L=2$ $M=64$ $S=2$ $R_c=2/3$ BICMB-OFDM-SG with different precoding selections and without precoding.}
\label{fig:pre_2L2S0.67R2x2}
\end{figure}
\else
\begin{figure}[!t]
\centering \includegraphics[width = 1.0\linewidth]{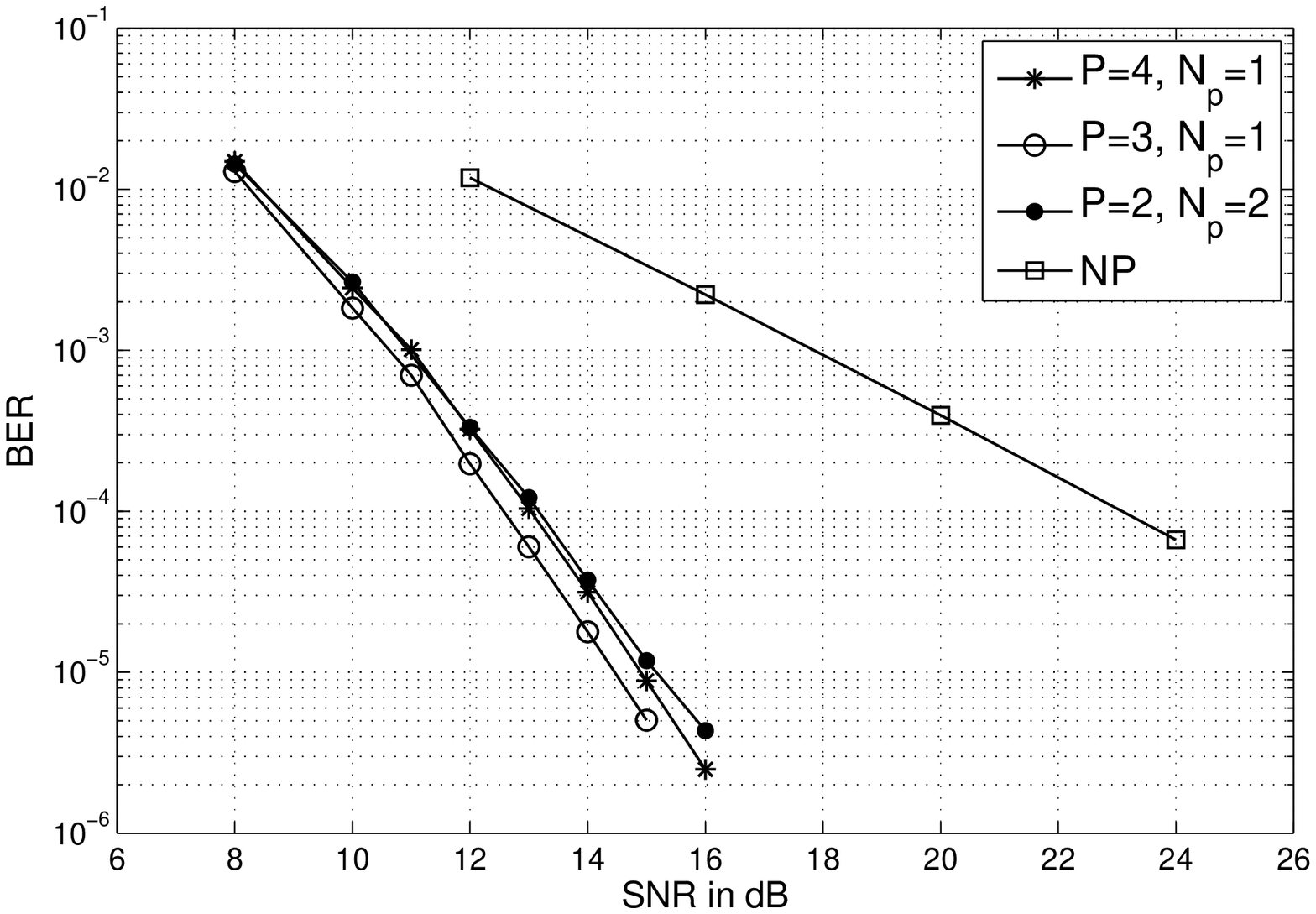}
\caption{BER vs. SNR for $2 \times 2$ $L=2$ $M=64$ $S=2$ $R_c=2/3$ BICMB-OFDM-SG with different precoding selections and without precoding.}
\label{fig:pre_2L2S0.67R2x2}
\end{figure}
\fi

Fig. \ref{fig:pre_2L2S0.67R2x2} shows the BER performance of $2 \times 2$ $L=2$ $M=64$ $S=2$ $R_c=2/3$ BICMB-OFDM-SG with precoding for different selections of $P$ and $N_p$ and without precoding. Since $R_cSL > 1$, full diversity cannot be achieved without precoding, and the diversity is $2$. On the other hand, both $P=4$ $N_p=1$ shown in Fig. \ref{fig:pre_2L2S2x2} and $P=3$ $N_p=1$ are effective precoding selections to provide the full diversity of $8$, while $P=2$ $N_p=2$ with diversity of $5$ cannot offer full diversity because $\lceil R_cSL \rceil = 3$. As discussed in Section \ref{subsec:pre_complexity}, $P=4$ has lower worst-case ML decoding complexity of $\mathcal{O}(N_m^{1.5})$ than $P=3$ of $\mathcal{O}(N_m^{2})$. However, $P=3$ achieves slightly better performance, which is less than $0.5\mathrm{dB}$, than $P=4$ in this case. Note that in the case of $P=2$ $N_p=2$, in order to achieve relatively high diversity, the first subchannel of the first subcarrier is precoded with the second subchannel of the second subcarrier, while the second subchannel of the first subcarrier is precoded with the first subchannel of the second subcarrier.

\ifCLASSOPTIONonecolumn
\begin{figure}[!t]
\centering \includegraphics[width = 1.0\linewidth]{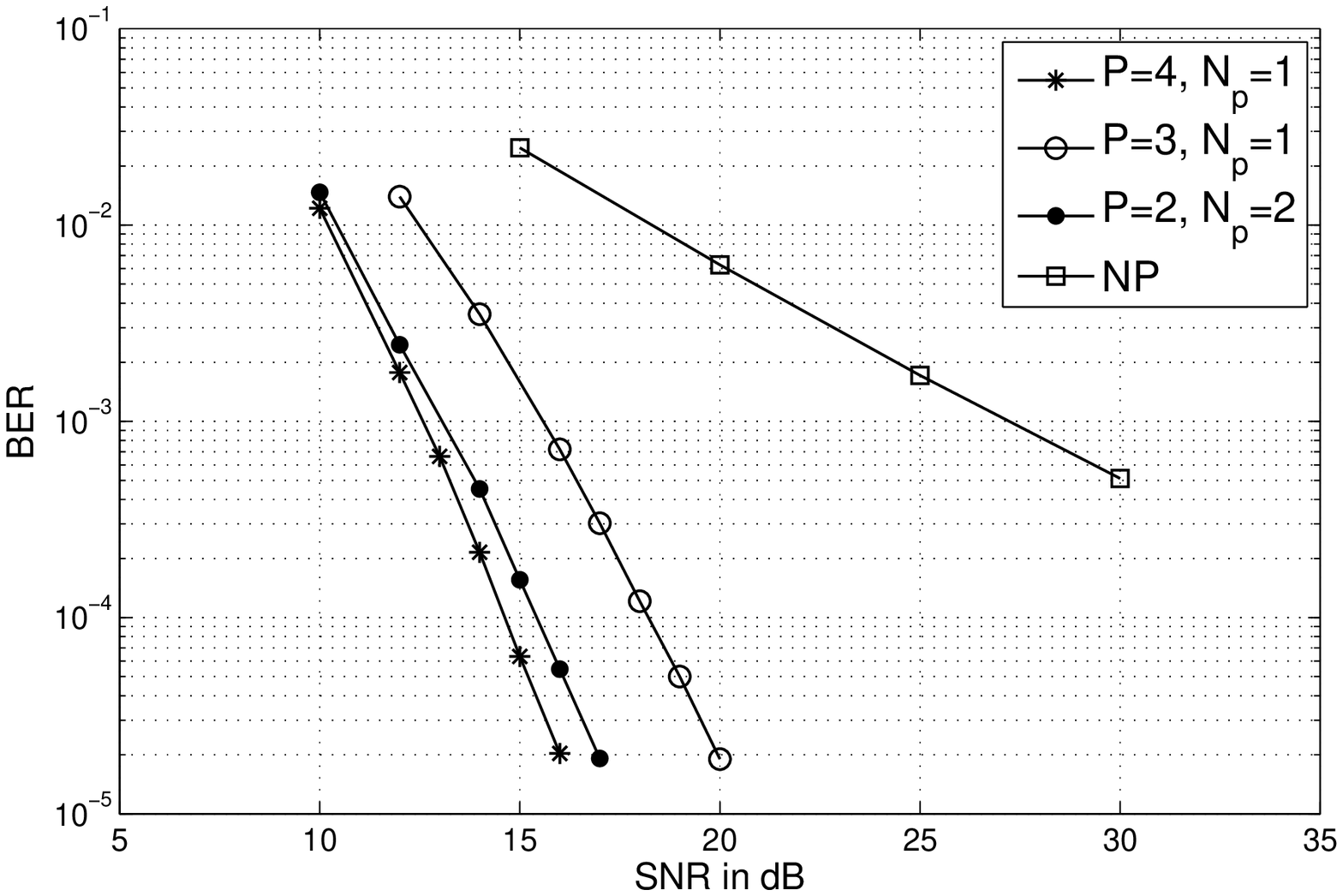}
\caption{BER vs. SNR for $2 \times 2$ $L=2$ $M=64$ $S=2$ $R_c=4/5$ BICMB-OFDM-SG with different precoding selections and without precoding.}
\label{fig:pre_2L2S0.8R2x2}
\end{figure}
\else
\begin{figure}[!t]
\centering \includegraphics[width = 1.0\linewidth]{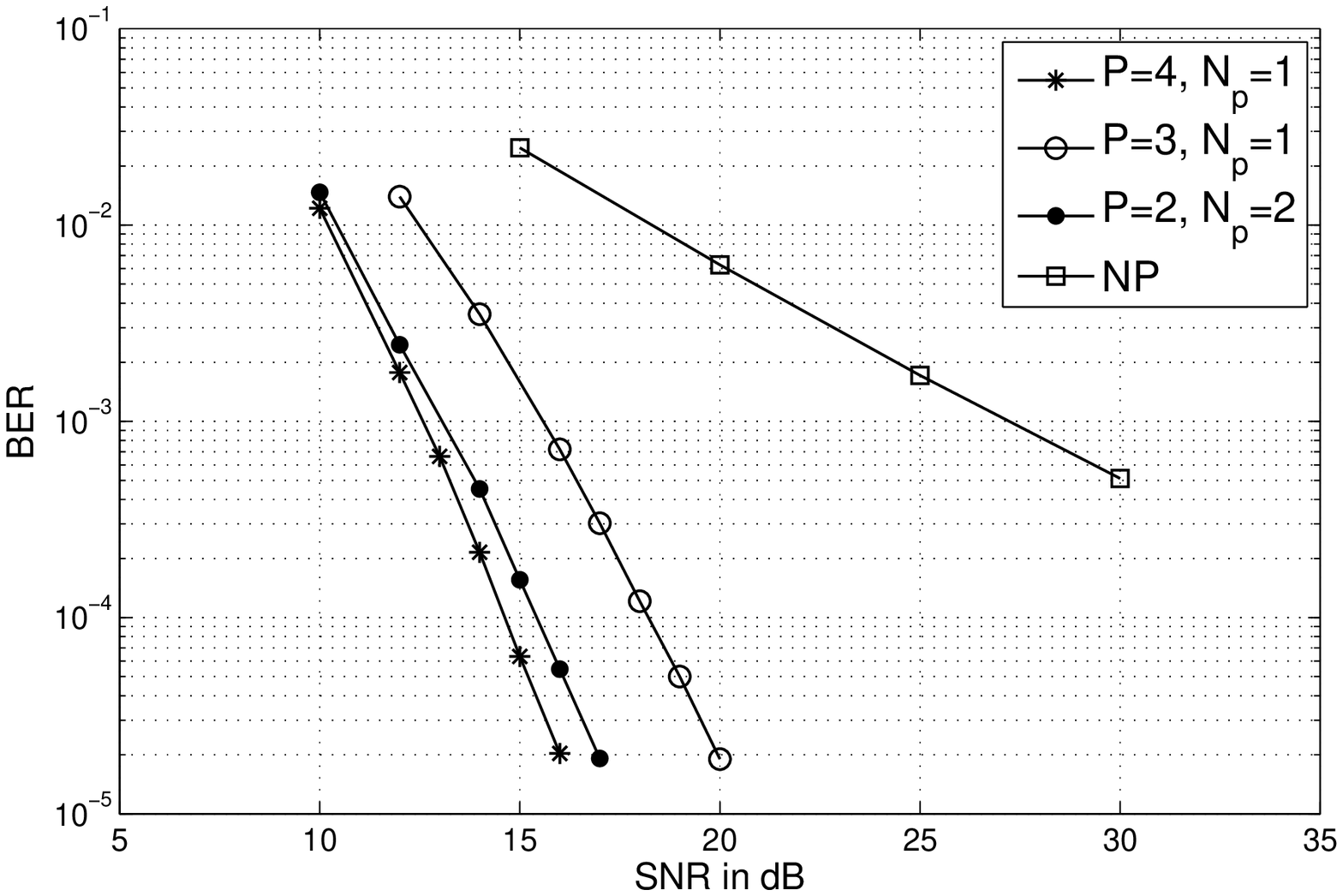}
\caption{BER vs. SNR for $2 \times 2$ $L=2$ $M=64$ $S=2$ $R_c=4/5$ BICMB-OFDM-SG with different precoding selections and without precoding.}
\label{fig:pre_2L2S0.8R2x2}
\end{figure}
\fi

Similarly, Fig. \ref{fig:pre_2L2S0.8R2x2} shows the BER performance of $2 \times 2$ $L=2$ $M=64$ $S=2$ $R_c=4/5$ BICMB-OFDM-SG with precoding for different selections of $P$ and $N_p$ and without precoding. Due to the fact that $R_cSL > 1$, full diversity cannot be provided without precoding and the diversity is $1$. On the other hand, $P=4$ $N_p=1$ shown in Fig. \ref{fig:pre_2L2S2x2} is an effective precoding selection to restore the full diversity of $8$, while $P=3$ $N_p=1$ and $P=2$ $N_p=2$ with diversity orders of $4$ and $5$ respectively cannot recover full diversity since $\lceil R_cSL \rceil = 4$. Note that in order to achieve relatively high diversity, in the case of $P=3$ $N_p=1$, the first subchannel of the second subcarrier is non-precoded, while for $P=2$ $N_p=2$, the first subchannel of the first subcarrier is precoded with the second subchannel of the second subcarrier, and the second subchannel of the first subcarrier is precoded with the first subchannel of the second subcarrier.

\subsection{$2 \times 2$ $L=4$ Precoded BICMB-OFDM-SG} \label{subsec:results_4taps_2x2}

\ifCLASSOPTIONonecolumn
\begin{figure}[!t]
\centering \includegraphics[width = 1.0\linewidth]{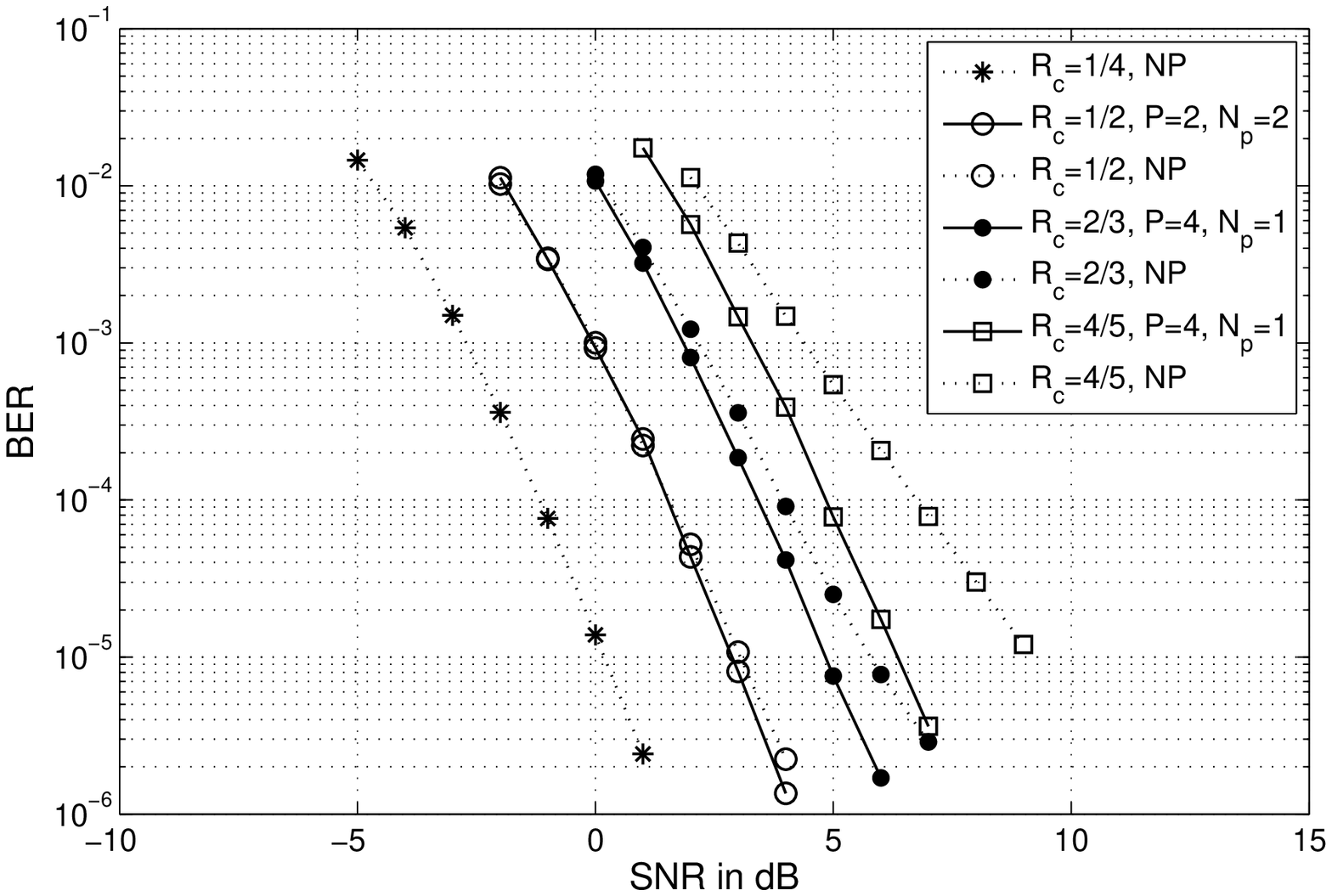}
\caption{BER vs. SNR for $2 \times 2$ $L=4$ $M=64$ $S=1$ BICMB-OFDM-SG with and without precoding.}
\label{fig:pre_4L1S2x2}
\end{figure}
\else
\begin{figure}[!t]
\centering \includegraphics[width = 1.0\linewidth]{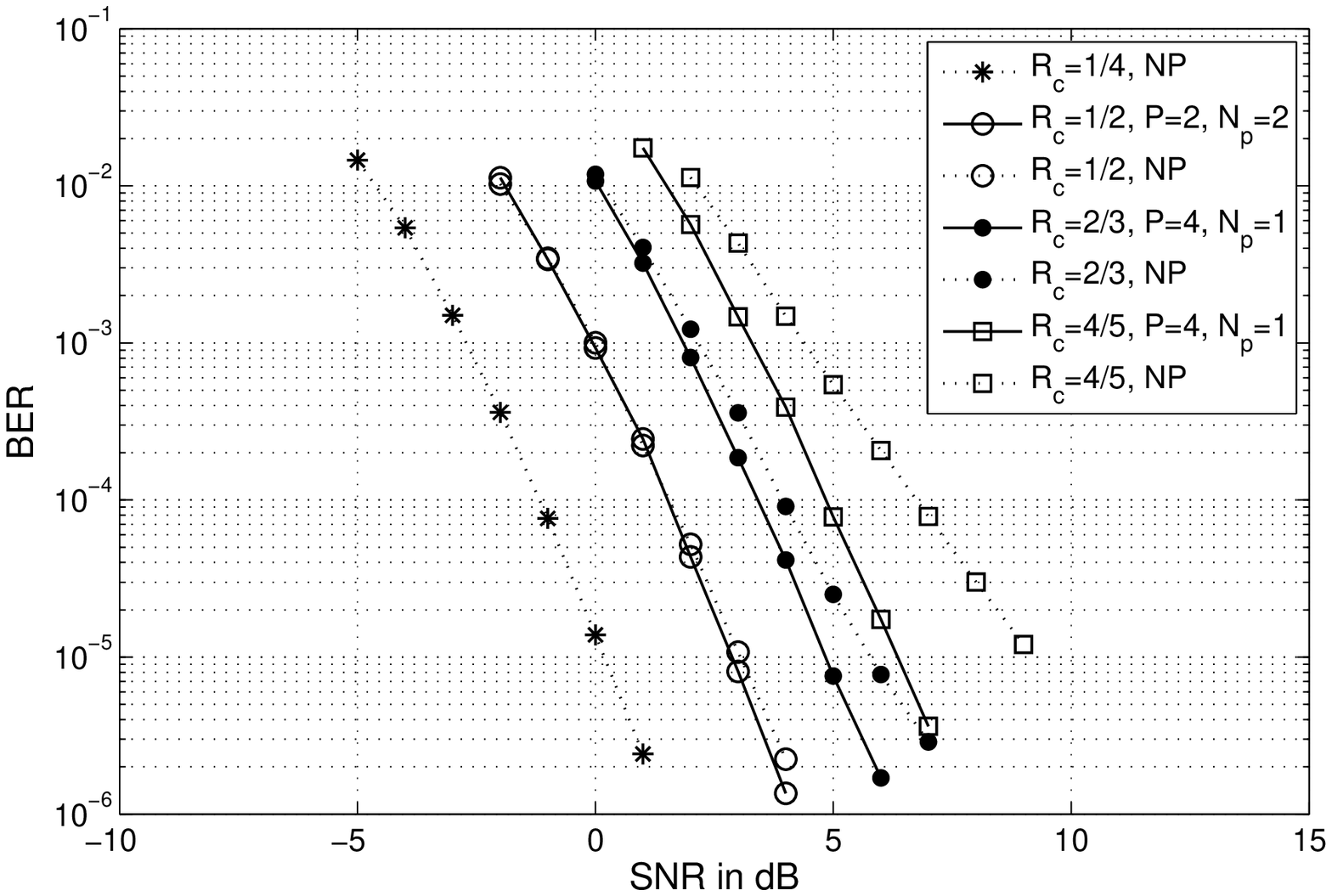}
\caption{BER vs. SNR for $2 \times 2$ $L=4$ $M=64$ $S=1$ BICMB-OFDM-SG with and without precoding.}
\label{fig:pre_4L1S2x2}
\end{figure}
\fi

Fig. \ref{fig:pre_4L1S2x2} shows the BER performance of $2 \times 2$ $L=4$ $M=64$ $S=1$ BICMB-OFDM-SG with and without precoding for different $R_c$. For $R_c=1/4$, full diversity of $16$ can be provided even without precoding as shown in Fig \ref{fig:np_div}, because $R_cSL \leq 1$ \cite{Li_DA_BICMB_OFDM}. On the other hand, in the cases of $R_c=1/2$, $R_c=2/3$, and $R_c=4/5$, the diversity orders are $12$, $8$, and $4$ respectively, and full diversity cannot be achieved since $R_cSL > 1$. However, full diversity of $16$ can be recovered by applying the full-diversity precoding design proposed in this paper. The corresponding selections are $P=2$ $N_p=2$, $P=4$ $N_p=1$, and $P=4$ $N_p=1$ for $R_c=1/2$, $R_c=2/3$, and $R_c=4/5$, respectively.

\ifCLASSOPTIONonecolumn
\begin{figure}[!t]
\centering \includegraphics[width = 1.0\linewidth]{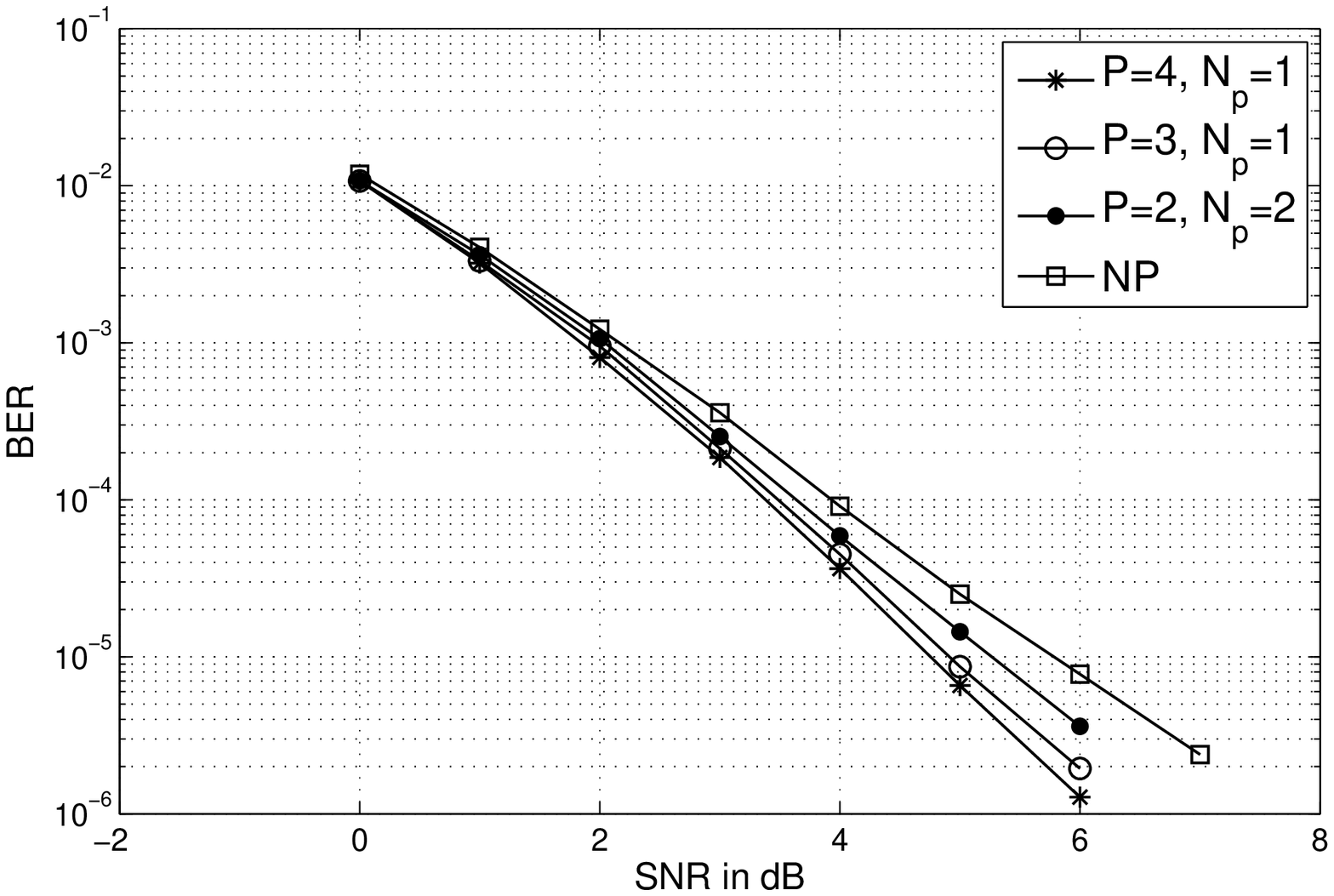}
\caption{BER vs. SNR for $2 \times 2$ $L=4$ $M=64$ $S=1$ $R_c=2/3$ BICMB-OFDM-SG with different precoding selections and without precoding.}
\label{fig:pre_4L1S0.67R2x2}
\end{figure}
\else
\begin{figure}[!t]
\centering \includegraphics[width = 1.0\linewidth]{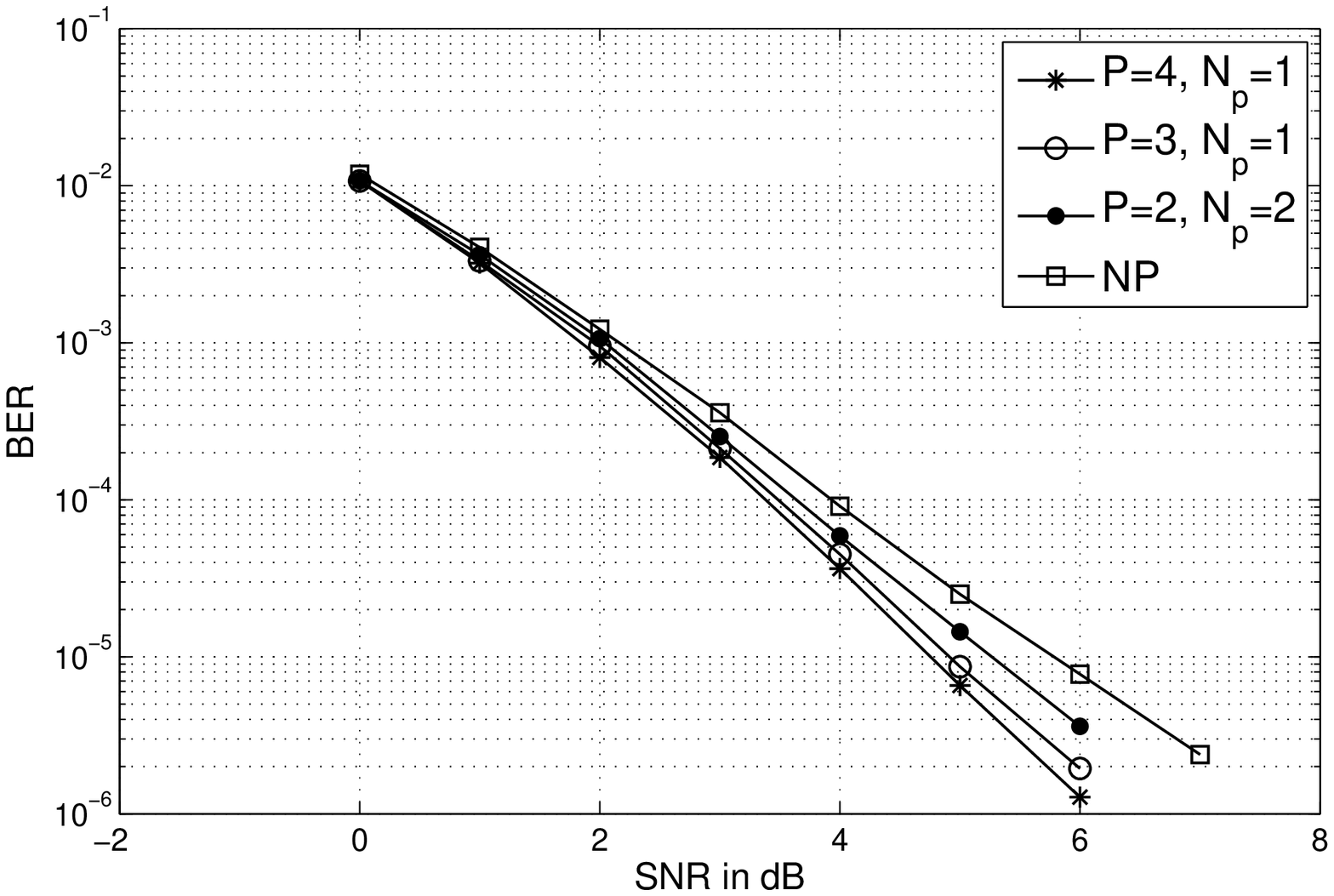}
\caption{BER vs. SNR for $2 \times 2$ $L=4$ $M=64$ $S=1$ $R_c=2/3$ BICMB-OFDM-SG with different precoding selections and without precoding.}
\label{fig:pre_4L1S0.67R2x2}
\end{figure}
\fi

Fig. \ref{fig:pre_4L1S0.67R2x2} shows the BER performance of $2 \times 2$ $L=4$ $M=64$ $S=1$ $R_c=2/3$ BICMB-OFDM-SG with precoding for different selections of $P$ and $N_p$ and without precoding. Because $R_cSL > 1$, full diversity cannot be achieved without precoding, and the diversity is $8$. On the other hand, $P=4$ $N_p=1$ shown in Fig. \ref{fig:pre_4L1S2x2} is an effective precoding selection to provide the full diversity of $16$, while $P=3$ $N_p=1$ and $P=2$ $N_p=2$ with diversity orders of $12$ and $8$ respectively cannot offer full diversity because $\lceil L/P \rceil P > SL$ and $\lceil R_cSL \rceil = 3$ respectively. Note that in order to achieve relatively high diversity, in the case of $P=3$ $N_p=1$, the subchannel of the second subcarrier is non-precoded, while for $P=2$ $N_p=2$, the subchannel of the first subcarrier is precoded with the subchannel of the third subcarrier, and the subchannel of the second subcarrier is precoded with the subchannel of the fourth subcarrier.

\ifCLASSOPTIONonecolumn
\begin{figure}[!t]
\centering \includegraphics[width = 1.0\linewidth]{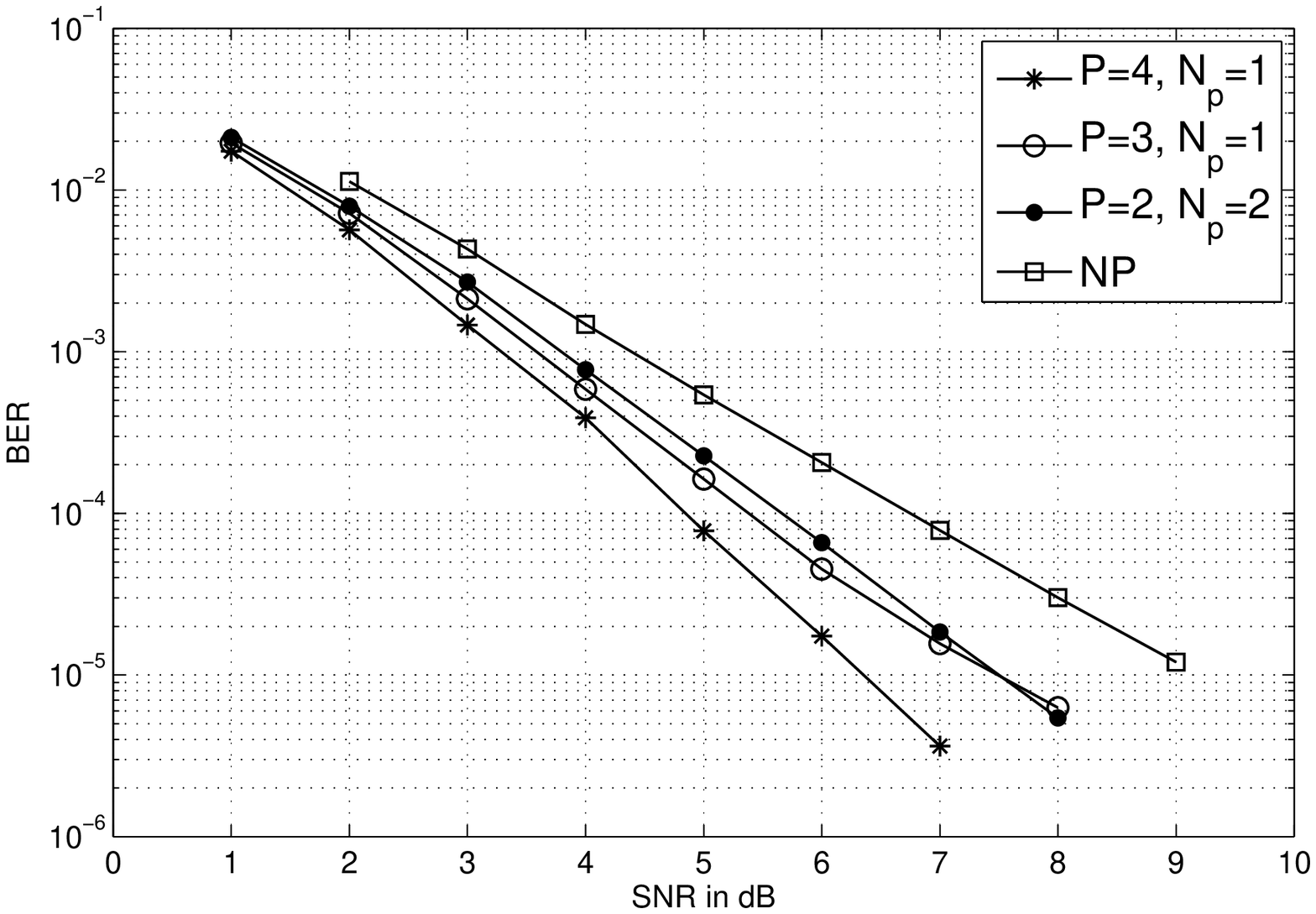}
\caption{BER vs. SNR for $2 \times 2$ $L=4$ $M=64$ $S=1$ $R_c=4/5$ BICMB-OFDM-SG with different precoding selections and without precoding.}
\label{fig:pre_4L1S0.8R2x2}
\end{figure}
\else
\begin{figure}[!t]
\centering \includegraphics[width = 1.0\linewidth]{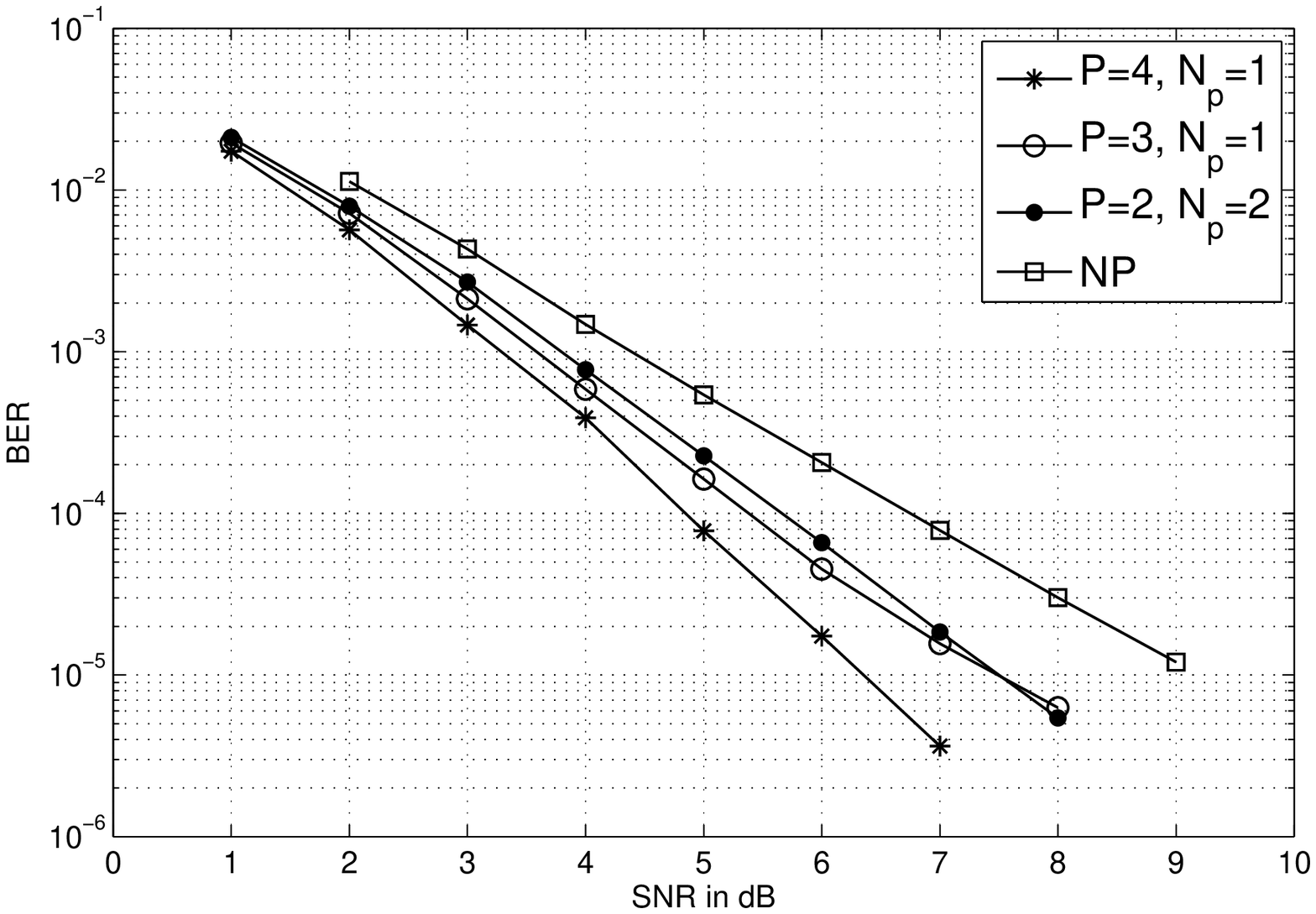}
\caption{BER vs. SNR for $2 \times 2$ $L=4$ $M=64$ $S=1$ $R_c=4/5$ BICMB-OFDM-SG with different precoding selections and without precoding.}
\label{fig:pre_4L1S0.8R2x2}
\end{figure}
\fi

Similarly, Fig. \ref{fig:pre_4L1S0.8R2x2} shows the BER performance of $2 \times 2$ $L=4$ $M=64$ $S=1$ $R_c=4/5$ BICMB-OFDM-SG with precoding for different selections of $P$ and $N_p$ and without precoding. Since $R_cSL > 1$, full diversity cannot be offered without precoding and the diversity is $4$. On the other hand, $P=4$ $N_p=1$ shown in Fig. \ref{fig:pre_4L1S2x2} is an effective precoding selection to recover the full diversity of $16$ while $P=3$ $N_p=1$ and $P=2$ $N_p=2$ with diversity orders $4$ and $8$ respectively cannot restore full diversity because $\lceil R_cSL \rceil = 4$. Note that in order to achieve relatively high diversity, in the case of $P=3$ $N_p=1$, the subchannel of the second subcarrier is non-precoded, while for $P=2$ $N_p=2$, the subchannel of the first subcarrier is precoded with the subchannel of the third subcarrier, and the subchannel of the second subcarrier is precoded with the subchannel of the fourth subcarrier.

\subsection{$4 \times 4$ $L=2$ Precoded BICMB-OFDM-SG} \label{subsec:results_2taps_4x4}
\ifCLASSOPTIONonecolumn
\begin{figure}[!t]
\centering \includegraphics[width = 1.0\linewidth]{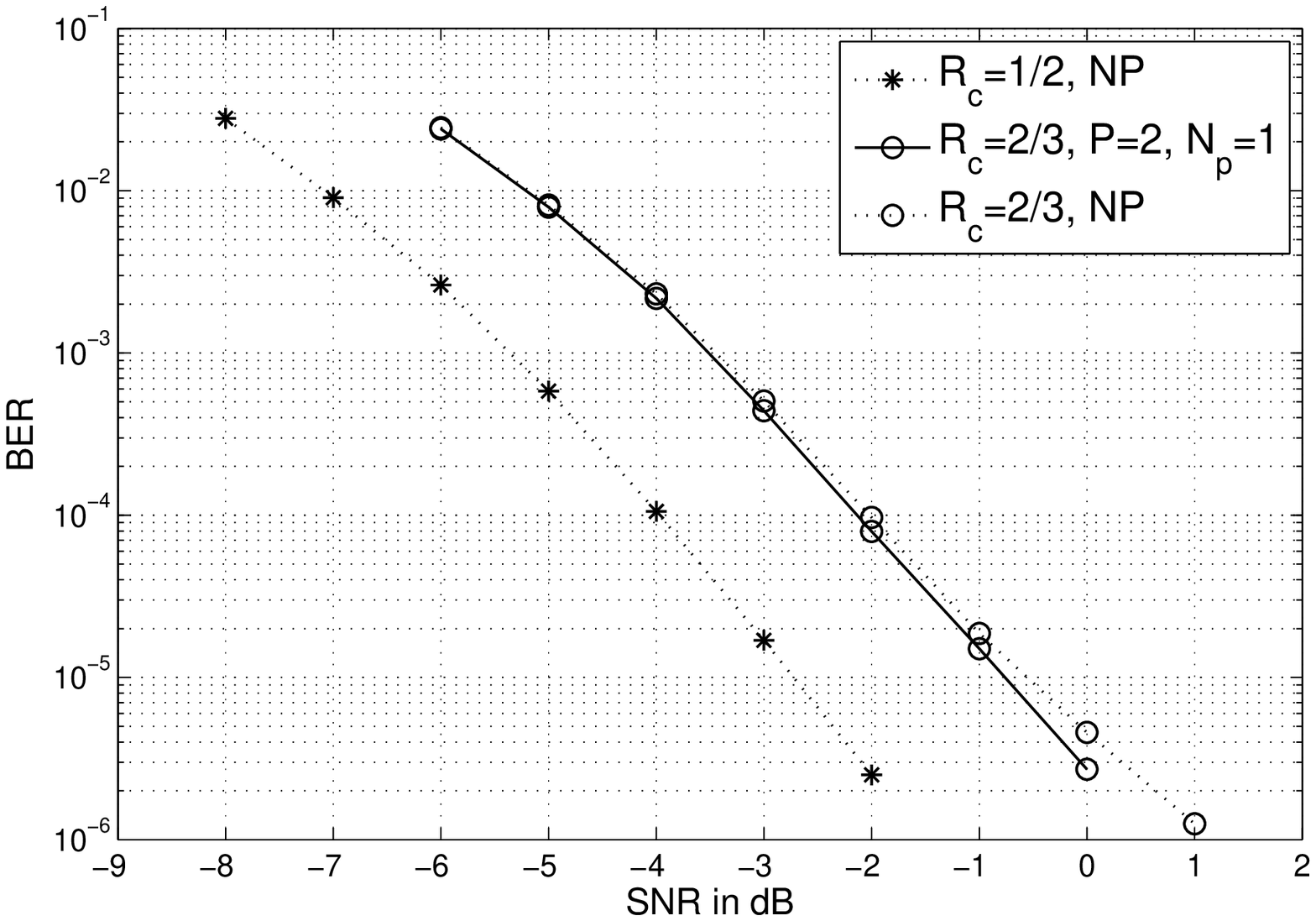}
\caption{BER vs. SNR for $4 \times 4$ $L=2$ $M=64$ $S=1$ BICMB-OFDM-SG with and without precoding.}
\label{fig:pre_2L1S4x4}
\end{figure}
\else
\begin{figure}[!t]
\centering \includegraphics[width = 1.0\linewidth]{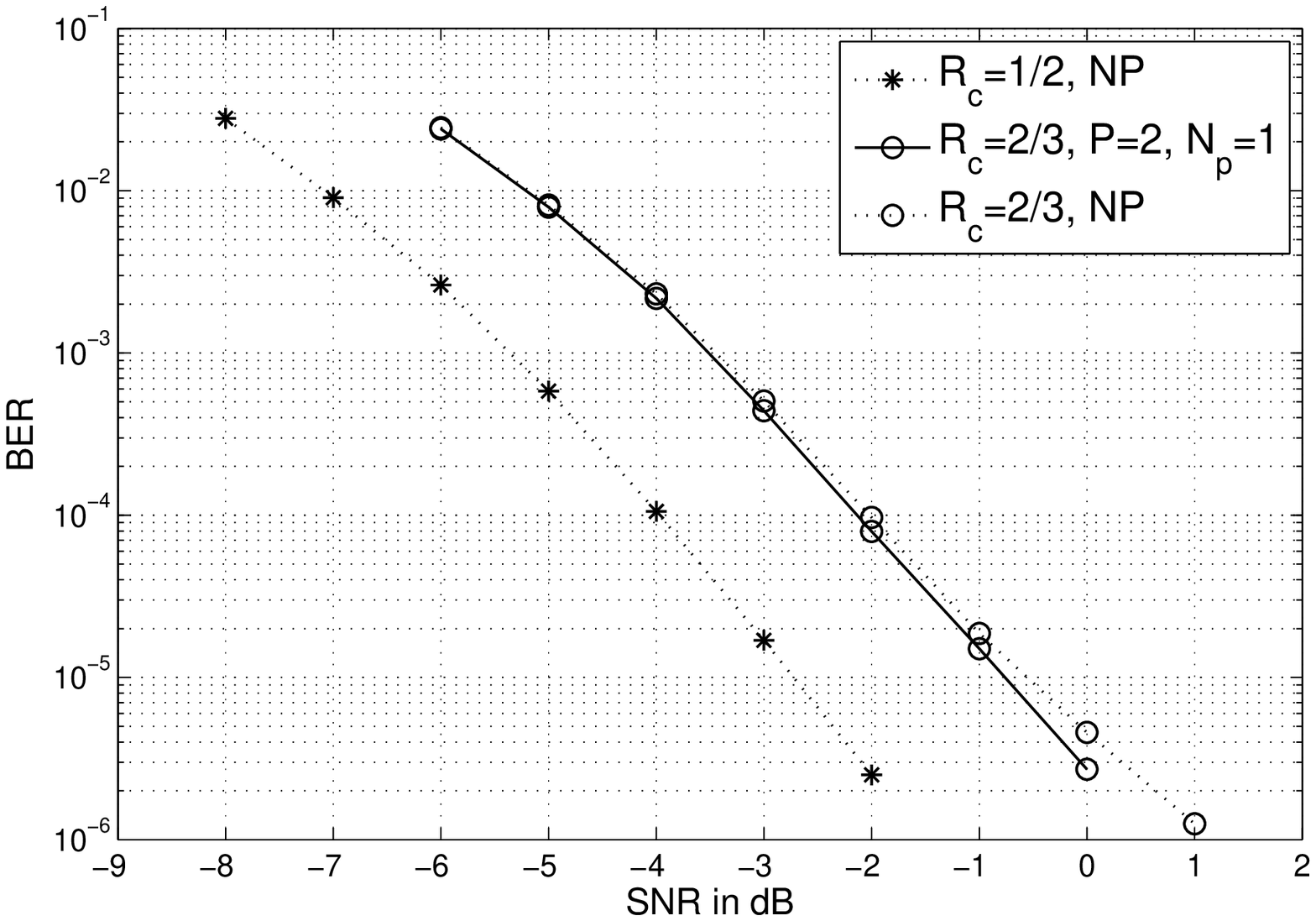}
\caption{BER vs. SNR for $4 \times 4$ $L=2$ $M=64$ $S=1$ BICMB-OFDM-SG with and without precoding.}
\label{fig:pre_2L1S4x4}
\end{figure}
\fi

Fig. \ref{fig:pre_2L1S4x4} shows the BER performance of $4 \times 4$ $L=2$ $M=64$ $S=1$ BICMB-OFDM-SG with and without precoding for different $R_c$. For $R_c=1/2$, full diversity order of $32$ can be achieved even if precoding is not applied as shown in Fig \ref{fig:np_div}, since $R_cSL \leq 1$ \cite{Li_DA_BICMB_OFDM}. On the other hand, in the case of $R_c=2/3$, the diversity order is $16$ instead of full diversity because $R_cSL > 1$. Nevertheless, full diversity of $32$ is recovered with the full-diversity precoding design proposed in this paper of $P=2$ and $N_p=1$. 

\subsection{Discussion} \label{subsec:result_discussion}
For a BER versus SNR curve, the negative diversity value is the slope while the coding gain \cite{Jafarkhani_STC} and array gain \cite{Andersen_AGC} reflect the position. As a result, although larger diversity provides better performance, best coding gain and array gain with full diversity can provide optimal performance. In this paper, the focus is drawn on achieving full diversity in terms of performance for BICMB-OFDM-SG without specific concentration on coding gain and array gain. In addition, equal power for each transmit antenna is assumed in this paper. Unequal power distribution with corresponding precoding design to achieve optimal performance is considered as future works. 

In the figures of this section, some curves of BICMB-OFDM-SG achieve full diversity but with different gains. The performance differences result from different reasons. In Fig. \ref{fig:np_div}, both $R_c=1/2$ $S=1$ and $R_c=1/4$ $S=2$ for $2\times2$ $L=2$ non-precoded BICMB-OFDM-SG achieve full diversity of $8$ with the same bit data rate. The performance disadvantage of $R_c=1/4$ $S=2$ is caused mainly by the usage of subchannels without only the largest eigenvalues for each subcarrier. 
Note that the reference curves are for flat fading MIMO channels and the bit data rates are much less than BICMB-OFDM-SG. In Fig. \ref{fig:pre_2L1S2x2}, Fig. \ref{fig:pre_2L2S2x2}, Fig. \ref{fig:pre_4L1S2x2}, and Fig. \ref{fig:pre_2L1S4x4}, the performance differences of full diversity curves result from the different employed convolutional codes, as a convolutional code with higher rate provides worse performance \cite{Lin_ECC} but greater bit data rate in general. Moreover, in Fig. \ref{fig:pre_2L2S0.67R2x2}, both $P=4$ $N_p=1$ and $P=3$ $N_p=1$ are effective precoding selections to achieve the full diversity of $8$ for $2 \times 2$ $L=2$ $M=64$ $S=2$ $R_c=2/3$ BICMB-OFDM-SG. The performance difference is caused by the different coding gains of the employed precoding techniques. 

As presented in Section \ref{sec:basis}, BICMB-OFDM-SG requires the knowledge of CSI at the Transmitter (CSIT), which is usually partial and imperfect in practice due to the bandwidth limitation and the channel estimation errors, respectively. Recently, limited CSIT feedback techniques have been introduced to achieve a performance close to the perfect CSIT case for both uncoded and coded SVD-based beamforming systems \cite{Narula_EUSI, Xia_TB_LRF, Love_LF_SM, Sengul_BICMB_ICSIT}. For these techniques, a codebook of precoding matrices is known both at the transmitter and receiver. The receiver selects the precoding matrix that satisfies a desired criterion, and only the index of the precoding matrix is sent back to the transmitter. In practice, similar techniques can be applied to BICMB-OFDM-SG.

As discussed in Section \ref{sec:diversity}, the PEPs with the worst diversity order dominate the overall performance. For BICMB-OFDM-SG without the condition $R_cSL \leq 1$, the PEPs without full diversity can be improved by applying the full-diversity precoding design proposed in this paper so that full diversity is achieved. If the diversity without full-diversity precoding is relatively small compared to full diversity, substantial improvement can be achieved by the full-diversity precoding design, e.g., the cases of $R_c=2/3$ and $R_c=4/5$ in Fig. \ref{fig:pre_2L2S2x2}. On the other hand, if the diversity without full-diversity precoding is close to full diversity, the advantage of the full-diversity precoding design may start at the SNR providing very low BER. In that case, its value depends on the BERs of different applications. Take the case of $R_c=1/2$ in Fig. \ref{fig:pre_4L1S2x2} as an example, if the BER requirement is $10^{-5}$, then precoding may not be necessary. However, if the BER requirement is $10^{-9}$, precoding may be worthwhile.

With the full-diversity precoding design proposed in this paper, more choices of BICMB-OFDM-SG with different trade-offs among performance, transmission rate, and decoding complexity are provided. Take Fig. \ref{fig:pre_2L2S2x2} as an example, without precoding, the case of $R_c=1/4$ achieves full diversity. However, increasing the transmission rate by employing convolutional codes with higher rates $R_c=1/2$, $R_c=2/3$, and $R_c=4/5$ results in the loss in performance. By applying the proposed full-diversity precoding design, full diversity can be recovered which trades off with higher decoding complexity. Moreover, higher rates of  $R_c=2/3$ and $R_c=4/5$ cause more increased decoding complexity than the case of $R_c=1/2$. The most proper choice varies which depends on the different requirements on performance, throughput, and decoding complexity.

\section{Conclusions} \label{sec:conclusions}

In this paper, a full-diversity precoding design is developed for BICMB-OFDM-SG without the full diversity restriction of $R_cSL \leq 1$. The design provides a sufficient method to guarantee full diversity while minimizing the increased decoding complexity caused by precoding. With this method, more choices are offered with different trade-offs among performance, transmission rate, and decoding complexity. As a result, BICMB-OFDM-SG becomes a more flexible broadband wireless communication technique.

\section*{Acknowledgment}
The authors would like to thank the anonymous reviewers whose valuable comments improved the quality of the paper. 


\bibliographystyle{IEEEtran}
\bibliography{IEEEabrv,Mybib}

\end{document}